\documentclass[fleqn,10pt]{wlscirep}
\usepackage[utf8]{inputenc}
\usepackage[T1]{fontenc}
\usepackage{multirow}
\usepackage[section]{placeins}
\title{Space Object Identification and Classification from
Hyperspectral Material Analysis}

\author[1,*]{Massimiliano Vasile}
\author[1]{Lewis Walker}
\author[2]{Andrew Campbell}
\author[1]{Simão Marto}
\author[2]{Paul Murray}
\author[2]{Stephen Marshall}
\author[3]{Vasili Savitski}
\affil[1]{University of Strathclyde, Mechanical and Aerospace Engineering, James Weir Building, 75 Montrose Street, Glasgow, G1 1XJ}
\affil[2]{University of Strathclyde, Electronic and Electrical Engineering, James Weir Building, 75 Montrose Street, Glasgow, G1 1XJ}
\affil[3]{Fraunhofer Centre for Applied Photonics, Level 5 Technology and Innovation Centre, 99 George Street, Glasgow, United Kingdom, G1 1RD}

\affil[*]{massimiliano.vasile@strath.ac.uk}

\begin{abstract}
This paper presents a data processing pipeline designed to extract information from the hyperspectral signature of unknown space objects. The methodology proposed in this paper determines the material composition of space objects from single pixel images. Two techniques are used for material identification and classification: one based on machine learning and the other based on a least square match with a library of known spectra. From this information, 
 a supervised machine learning algorithm is used to classify the object into one of several categories based on the detection of materials on the object. The behaviour of the material classification methods is investigated under non-ideal circumstances, to determine the effect of weathered materials, and the behaviour when the training library is missing a material that is present in the object being observed. Finally the paper will present some preliminary results on the identification and classification of space objects.

\end{abstract}
\begin{document}

\flushbottom
\maketitle
%
%
\thispagestyle{empty}

\section{Introduction}
The composition of space objects can be characterised based on the spectra of light reflecting off these objects, as each material present will commonly give unique light spectra. In the last 20 years, various techniques based on spectroscopy, spectral analysis and colour photometry have been proposed  \cite{JORGENSEN20041021,Abercromby2005,Reyes2018CharacterizationOS,CARDONA2016514,zhao2016multicolor,VANANTI20172488,WILLISON20161318}. However, the work presented in these references is mainly observational. Spectra were observed for known sets of objects and these spectra were then associated to known material types. 

Colour photometry \cite{zhao2016multicolor} has been proposed to differentiate between different objects but was not used to develop a classification system.
Meanwhile techniques using colour indexes, alongside characteristics of the spectra, have been used to develop a classification system, see for example \cite{CARDONA2016514,VANANTI20172488}.
Both colour indexing and photometry produce relatively sparse spectral data in comparison to techniques such as hyperspectral imaging and spectrometry. In recent years, both of these spectrally rich techniques have proven useful in the classification and characterisation of asteroids \cite{Lind2021}. Multispectral imaging has also proven useful for close proximity navigation \cite{Esposito2018} but mainly using the near infrared part of the spectrum and for Earth Observation and remote sensing application, such as the identification of targets on the ground in \cite{Poojary2015}.

Classification of space objects has also been achieved using temporal rather than spectral information using light curve analysis \cite{wetterer2009,dianetti2018,Matsushita2019,chote2019,santoni2013,YANAGISAWA2012136}. Most recently, Machine Learning (ML) was introduced to classify objects directly from light curves \cite{Kerr2021,McNally2021}. As no spectral information is present these techniques rely on temporal information to reconstruct the attitude motion or the shape of space objects. As such while classification of known objects is possible, no information on the composition of unknown objects is obtained.

This paper will present methods for identifying space objects from time-varying hyperspectral sensor data using a pipeline that begins by identifying the materials present and then builds an understanding of the object before reaching a final classification. Specifically, it is proposed to first use spectral decomposition techniques to identify the abundance of each material present on the visible surface of the material at each point in time. From this prediction a probability score indicating the confidence of a significant abundance of material being present is computed. This information is then used to classify what type of satellite or other space object is observed. The proposed pipeline provides an explainable model that, even in the case of an unknown space object, would return useful information about its material composition. If the unknown object meets the expectations and assumptions of a known class, it will be labelled as belonging to that class. Otherwise the proposed pipeline allows unknown objects that do not fit any known class well to be categorised as an Unidentified Flying Object (UFO). 

 This paper builds on the authors original ideas presented in \cite{vasile2022intelligent, vasile2022hyperclass} and uses simulated data based on object and sensor models developed in the previous work. The main contribution of this paper is twofold: a) we improve the material identification approach previously presented in \cite{vasile2022hyperclass} including a comparison between a machine learning and a non-machine learning based identification approach, the ageing of the materials and a technique to account for unknown materials; b) we propose a probabilistic approach to the identification of materials and classification of space objects.

 We will show how, starting from high fidelity simulated data using real spectra acquired in lab environment, we can reliably unmix the spectra and identify the materials. Then we will show how objects can be classify with high probability from the expected combination of observable materials.

 The paper is structured as follows. In Section \ref{sec:pipeline} we provide an overview of the methodology, in Section \ref{sec:simulations} we recall the way we generate and process the hyperspectral data, this section is largely based on the work done in \cite{vasile2022intelligent}. In Section \ref{sec:materials} we present both the machine learning and the non-machine learning approaches to decompose the spectra and identify the materials. Finally in Section \ref{sec:classification} we present the classification system.

\section{Overview of the Object Identification and Classification Pipeline}
\label{sec:pipeline}
An overview of the data processing pipeline presented in this paper will first be given in this section before presenting each step in more detail in later sections.
A dataset of time-series spectral data of space-objects is required for this study. Ideally, this would come from real objects in orbit, but practically this presents several challenges. Firstly, in order to confirm that the models generalise well, the diversity of this dataset must be large. This means not only many different objects should be present in the dataset, but the training set must also contain a large range of orbital scenarios, trajectories relative to the observer, and illumination conditions even for the same object. Secondly, accurate evaluation necessitates ground truth data for those values which the models are to predict. In this context, this means having precise knowledge of the geometry, material distribution, and attitude motion at all times. Although the former two are theoretically possible to obtain, generally the materials used and their distributions over the surface of satellites is not something that is reported in literature. Additionally, if one wishes to include uncontrolled debris objects (eg spent rocket bodies) in the dataset, precise knowledge of the attitude will not be obtainable for real data. For these reasons it was decided to construct numerical physics simulation software to produce training data for the ML models.

The simulation models, as will be detailed in Section \ref{sec:simulations}, propagate the orbital and attitude motion of arbitrary space objects, and simulate the sensor output (aggregated onto a single pixel) at each timestep. This produces a time-varying spectral response as the object moves and rotates relative to the observer, and the illumination conditions change. The simulation model can simulate either a ground-based sensor (accounting for wavelength-dependent atmospheric extinction and Earth rotation), or a space-based sensor which follows its own trajectory. 

\begin{figure}[ht]
\centering
\includegraphics[width=0.95\linewidth]{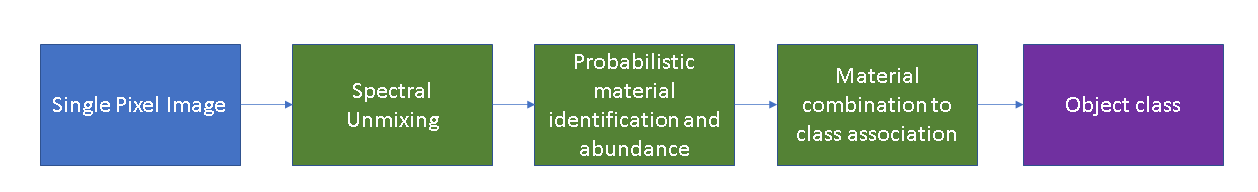}
\caption{Classification pipeline and probabilistic material identification.}
\label{fig:pipeline}
\end{figure}

The output of the sensor simulation is then decomposed into the proportion of signal due to each of the present materials. This can be done using machine learning or a more traditional constrained least squares approach, where both methods make use of a spectral library to determine the instantaneous material breakdown of the signal.The probability of a set of known materials being present on the object is then computed. 
The final step is to pass a detected combination of materials to a machine learning model trained to associate a given combination of materials to a class. This final step yields a probabilistic classification coming from the probability of detecting a combination of materials. A simple flow diagram of the proposed pipeline can be seen in Figure \ref{fig:pipeline}.

This staged architecture was developed in favour of a direct classification to produce a final pipeline with greater transparency of the classification process. Not only does this pipeline enable a logical flow to be followed through the data processing, but also outputs useful information at the intermediate steps, such as the material abundances.

\section{Simulation Models and Data Pre-processing}\label{sec:simulations}
Test data in this work was produced using a high-fidelity simulation model, developed by the authors in previous work \cite{vasile2022hyperclass,vasile2022intelligent}. Since the focus of this paper is the data processing pipeline, only a brief overview of the simulation framework will be given in this Section; for more details the authors refer the reader to their previous works.

The model takes as its input a 3D model representation of a space object to be simulated along with an initial state describing the osculating orbital elements, attitude, instantaneous angular velocity and inertia tensor. The model then propagates attitude and orbital motion, and at each timestep calculates the illumination conditions of each facet of the 3D model. Before input to the simulation model, the facets are each assigned a particular composition of materials, and a composite reflectance spectrum is constructed from the corresponding linear combination of materials. This, along with the instantaneous illumination conditions (illumination direction and view direction) are used to determine the visible and shadowed facets at each timestep, and aggregate the received light from all visible facets according to the Lambertian reflectance model. Reflectance and emission spectra are taken from lab experiments of known materials. Not only were these lab experiments used to gather the spectra required but also to investigate the consistency of the measured reflected spectra for different angles of incident light. 

\begin{figure}[ht]
\centering
\includegraphics[width=0.75\linewidth]{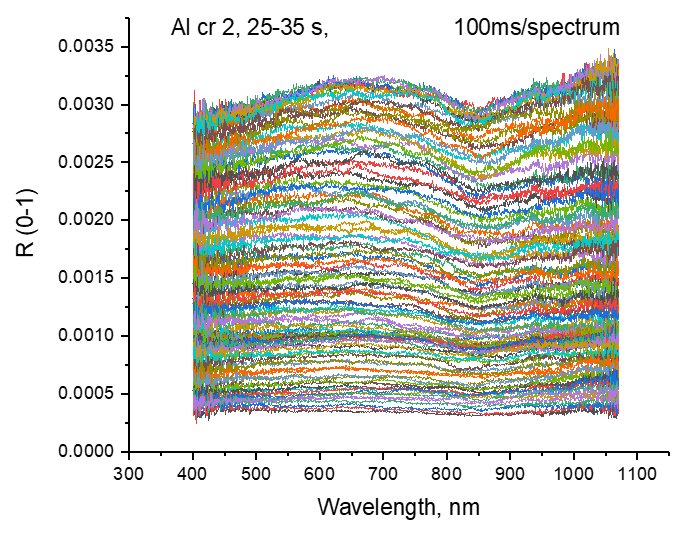}
\caption{Plot of multiple spectral measurements of the same aluminium foil at different orientations}
\label{fig:specConsistency}
\end{figure}

Spectra were acquired from laboratory experiments as follows.  
Materials were affixed to a cube placed on a rotation stage and many spectra were gathered while the cube was rotated to constantly change the relative angle between the face and the light source. While the total amount of light hitting the surface will inevitable change the overall amplitude of the gathered spectra we observed consistent spectra responses in terms of the location and relative magnitude of spectral features, as illustrated by the example for an aluminium foil shown in \ref{fig:specConsistency}. This gives us confidence that the spectra we use within the model and the simulation method described above is consistent with a real world scenario.

\begin{figure}[ht]
\centering
\includegraphics[width=0.95\linewidth]{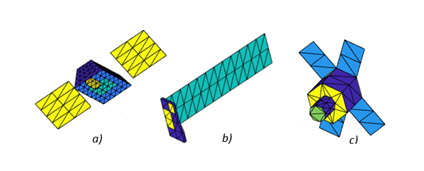}
\caption{Example of three simulated satellites: a) Iridium-NEXT, b) Starlink, c) DubaiSat-2. Colours correspond to regions with different materials. The reflectivity spectra for each region were obtained from laboratory experiments using mock-ups.}
\label{fig:example_space_obj}
\end{figure}

Figure \ref{fig:example_space_obj} shows an example of thee types of simulated space objects. The actual spectra coming from the laboratory experiments were placed on each of the triangles composing the surface of the object in Figure \ref{fig:example_space_obj}. Light intensity derived from the simulated illumination conditions in space was then added to spectra on each triangle.

For the sake of the tests in this paper all the objects where placed in the same Low Earth Orbit and the sensor was placed on the surface of the Earth. 
Orbits of the imaged objects were randomized to increase training data diversity. The training set is generated for one fixed location of the telescope on the surface of the Earth, defined by its latitude and longitude. In deployment, images are likely to be taken during twilight hours to maximize the reflected light from the object while minimizing atmospheric scatter and sunlight leakage into the telescope aperture. Hence for all simulations the telescope was initialised on the western terminator, i.e. observing objects in the early morning, coming over the horizon towards the sun. The orbital elements of the objects are generated randomly within the bounds in Table \ref{tab:OE_randomized}, which ensures that the objects begin their propagation between 25 and 10 degrees before zenith with respect to the observer. 5-minute observation arcs are simulated for all training samples, however the pipeline proposed is intentionally agnostic to the duration of the exposure and number of time samples.

\begin{table}[]
    \centering
    \begin{tabular}{c|c}
        Parameter (symbol) & Value range \\
        \hline 
        Semi-major axis ($a$) & [7471, 7671] km\\
        Eccentricity ($e$) & [0, 0.01]\\
        Inclination ($i$) & [-45, 45] $^o$ \\
        Argument of Periapsis ($\omega$) & [0, 360] $^o$\\
        RAAN ($\Omega$) & [-5, 5]$^o$ \\
        True Anomaly & 360$^o$ - $\omega$ - [10, 25]$^o$ \\
        Telescope latitude & [-5, 5] \\
        Telescope longitude & [-5, 5]
    \end{tabular}
    \caption{Range of initial orbital elements used for training data generation. The true anomaly first places the satellite at the ascending node $\theta = 360-\omega$, and then moves it backward in its orbit by 10 to 25 degrees}
    \label{tab:OE_randomized}
\end{table}

The initial orientation and angular velocity vectors are randomly initialised, with an upper bound on the rotation rate of 2 rotations per minute. Although many of the satellites simulated in this paper would in reality have nadir-pointing configurations, one possible application of this technology is locating lost satellites with which communication has been lost, and these targets may be tumbling. Taken together, the randomisation of all these parameters ensures a good level of diversity in the training set. For all simulations, an observation arc of 5 minutes was simulated, sampled at 1 Hz. The band width of the hyperspectral sensor was set to 5 nm. The simulation model is built to be compatible with arbitrary wavelength bands, but in these simulations we used the spectral range spanning 455 nm to 1035 nm, with a constant 5 nm band width. The incoming spectrum is simply integrated across each band's range to obtain the total collected power. This is then converted to sensor counts by dividing by the band's central photon energy. Shot noise and readout noise are accounted for in the sensor model.

The hyperspectral sensor was modelled as yielding no spatial information, thus in all experiments in this paper the light collected by the simulated sensor in a given wavelength band is captured by a single pixel. Thus, the received signal is two-dimensional, varying only in wavelength and time, as opposed to the more typical definition of hyperspectral imaging where there is a three-dimensional signal of two spatial and one spectral dimension.

A key pre-processing step for the simulated sensor output is the transformation from band-wise photon counts to a 
 hyperspectral colour-indexed representation of the signal, which is an application of colour indexing \cite{ZHAO20162269} to hyperspectral signals. This transformation removes distance and absolute brightness-related effects while preserving the spectral shape and relationships in brightness between different bands, which is the quantity of interest to this work. This transformation acts analogously to the normalising or scaling transformations that are typically applied to data before being passed to machine learning models for training or testing. The equation for the transformation is: 

\begin{equation}\label{eq:HRCI}
    \mathbf{s}(t,\lambda) \rightarrow -2.5\log_{10}\left(\frac{\mathbf{s}(t,\lambda)}{\mathbf{s}(t,\lambda_{ref})}\right)
\end{equation}
where $\mathbf{s}(t,\lambda)$ is the time-evolving spectrum received from the object, and $\lambda_{ref}$ is an arbitrary reference wavelength band.
For the simulation of the sensor on ground we include also the effect of the atmosphere.

With this model it is possible to define the attitude of the space objects for which the observation of is simulated\cite{vasile2022intelligent}. As the purpose of the proposed models is to predict the materials present from spectral data and what type of objects this may correspond to,  we elect to use randomised attitude for all objects rather than attempt to mimic the real world pointing directions of particular satellites. This was done to improve the robustness of our study as a greater number of observation scenarios could be investigated. This allows us to confirm if the proposed models can robustly identify materials on all faces of all objects simulated, not just those visible from a defined perspective. 
\section{Material Decomposition Methods}
\label{sec:materials}
In order to reconstruct the material-wise contribution to the received signal over time, two methods were investigated: one using machine learning and one using a more traditional, explainable method. These two methods are explained in this section and results of each will be presented. This step of the pipeline will allow us to extract the Material Abundance Curves (MAC) associated to each observed object, or curves providing, for each observation time step, the relative abundance of each material present in the received spectrum.

\subsection{Machine Learning Material Identification}
\label{sec:ml_materials}
We first approached the identification of individual materials in a spectrum by training an Artificial Neural Network (ANN) to decompose the spectrum.
In order to train an ANN to decompose a composite spectrum into its material components, a training set was constructed composed of $N$ random linear combinations of $m$ materials, constrained such that their sum is equal to 1. While it is an option to train the decomposition ANN using simulated sensor data from the simulation model, it was found that this approach effectively embeds the geometry of the underlying objects in the training set, reducing the generalisation performance when the model is used with unseen geometries. When instead training on these linear combinations of raw real spectra from laboratory experiments, a geometry-agnostic model is produced which performs similarly well across all geometries we exposed it to for validation.

The ANN developed for this stage had a relatively simple architecture - a fully-connected, feed-forward regression network, with two hidden layers and a combination of multiple activation functions on those hidden layers. The softmax activation function was applied to the output layer to ensure that the sum of all material contributions was equal to 1. The full list of hyperparameters in the final model can be seen in Table \ref{tab:ANN_hyperparams}. The model was trained with early stopping enabled such that the weights and biases revert to the best found set after 8 epochs of no improvement. The final model developed and used for later stages of the data processing was trained with a library of nine materials: black, white, red and green paints, aluminium foil, Multi Layer Insulation (MLI) (or gold colour) thermal blanket, copper and titanium metals, and GaAs solar panel. Paint reflectance spectra were obtained from the USGS spectral library \cite{clark2007usgs}, while the remaining spectra were obtained from our own laboratory measurements.

\begin{table}[ht]
\centering
\begin{tabular}{|l|l|}
\hline
Hyperparameter & Value  \\ \hline
\# hidden layers &  2 \\ \hline
\# hidden neurons &  200, 20 \\ \hline
Activation (h1) & ReLU \\ \hline
Activation (h2) & Sigmoid \\ \hline
Activation (output) & Softmax \\ \hline
Optimiser & Adam \\ \hline
Loss Function & MSE \\ \hline
Train set size & 8,100,000 samples \\ \hline
Test set size & 900,000 samples \\ \hline
Minibatch size & 100,000 samples\\ \hline
\end{tabular}
\caption{\label{tab:ANN_hyperparams}Hyperparameters of the decomposition ANN. The number of input and output neurons must be equal to the number of spectral bands and library materials respectively.}
\end{table}

\begin{figure}[ht]
\centering
\includegraphics[width=0.65\linewidth]{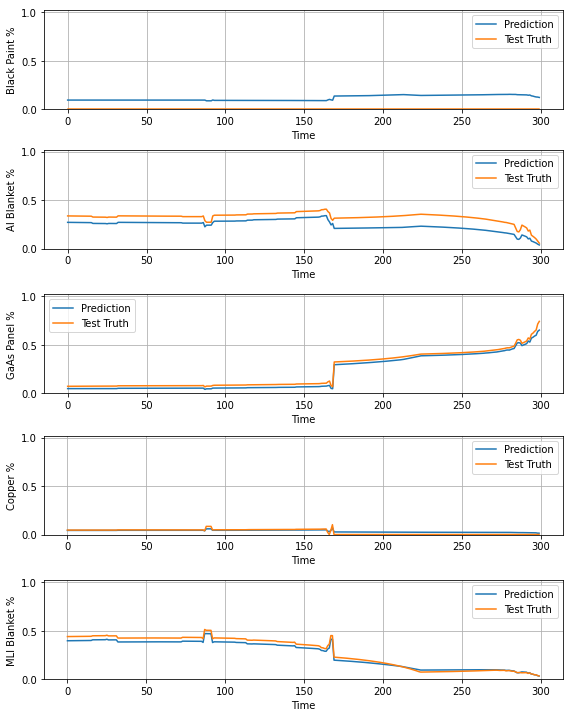}
\caption{Example of ANN predictions for material abundances for an Iridium-NEXT satellite. Over 200 simulations with differing orbital and rotational conditions, the mean squared error in the predictions was 5.32 $\times$ 10$^{-3}$ }
\label{fig:MACANN_iridium_example}
\end{figure}

After training on the dataset of spectral combinations, the ANN was capable of making accurate predictions of the material-wise breakdown of the received light for all tested geometries. Example curves for a simulation of an Iridium-NEXT satellite can be seen in Figure \ref{fig:MACANN_iridium_example}, showing a good fit to the ground truth with the exception of black paint, which was found to be generally overestimated, although, the overall shape of the curves is correctly predicted. Additionally, albeit the library typically contains more materials than are present in an individual test satellite, it did not predict the presence of materials that were not in the satellite being tested (for example, if a satellite does not contain white paint, the ANN did not predict any significant quantity of white paint despite it being in the library).

\subsection{Non-Machine Learning Material Identification}
\begin{figure}[ht]
\centering
\includegraphics[width=0.95\linewidth]{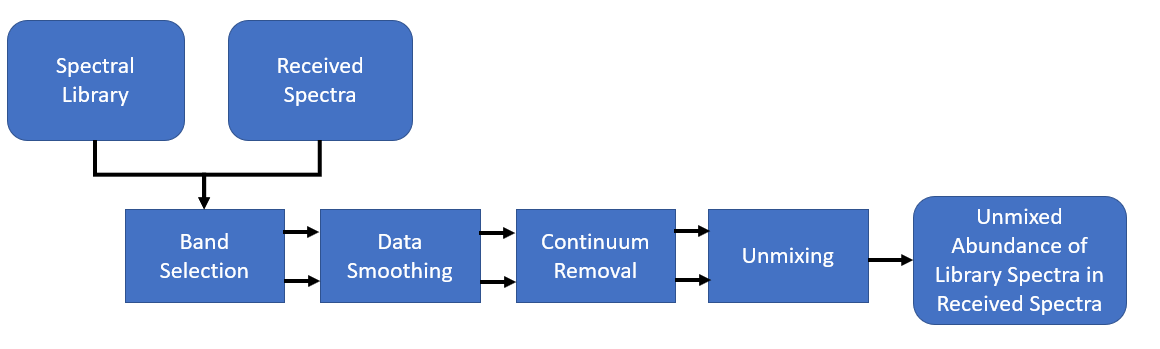}
\caption{Schematic of library unmixing algorithm}
\label{fig:schematic_LM}
\end{figure}

While good unmixing performance was achieved using machine learning techniques such methods often lack explainability and require time and data to re-train. Non-machine learning approaches were explored to ascertain if similar unmixing could be achieved in a more explainable way and improve confidence in the ability to unmix material abundances reliably from spectral data. 
A number of techniques exist in the literature for the unmixing of hyperspectral data. To successfully unmix a received signal two problems need to be solved; finding the fundamental spectra present in the data, i.e. the spectra of the materials that are present, and what proportion of these spectra make up the received spectra. The fundamental spectra that make up the received signals are more commonly referred to as endmember spectra and a wide range of techniques focus specifically on identifying these spectra. While the specific methods differ, these techniques typically assume that the endmember spectra will exist in relatively pure form (i.e. the spectrum is composed of near 100\% one material) in the data at some point in space or time. For example, well established solutions such as N-findR \cite{winter1999n} are very highly performance limited by the purity of received spectra. This technique assumes the volume of a simplex formed by the endmembers is larger than the volume of a simplex formed by any other combination of spectra. Using this assumption, the algorithm can search combination of spectra in the received data to find those that best represent the endmembers of the data. However, this means all returned endmembers are selected from the data, thus this method will only represent the purest spectra that exists in the data which may still be some combination of multiple materials. Methods other than the volume of a simplex can be used to compute the likelihood of a spectra being an endmember, such as orthogonal subspace projection \cite{harsanyi1994hyperspectral} or vertex component analysis \cite{nascimento2005vertex} among others \cite{heylen2011fully}, but the same limit will apply while endmembers are selected from the received spectra. Alternative approaches exist using methods such as SVD \cite{zhang2016exact} or Tucker Decomposition \cite{nie2021space}, which have recently been applied successfully to unmix materials in hyperspectral images of scaled down satellites in a laboratory environment. However, in \cite{nie2021space} the authors note that the higher the proportion of pure pixels the better the more accurate the resulting endmember spectra will be. The test data used in the worst case still has 25\% unmixed pixels in the data. Such methods are thus not best suited for single pixel spectral measurements where the level of mixing is likely to always be high.
To resolve these challenges, it is proposed to unmix the received spectra based on a library of candidate materials that are likely to exist in space. This approach is conceptually similar to the ANN proposed in Section 4.1.  The idea is to compute, for each point in the time-series, the weighted linear combination of reference spectra in the library that would produce the received spectra. If successful this returns the abundance of each material present in each received spectra, with materials that are not present returning a 0\% contribution. A schematic of the proposed library unmixing method is shown in Figure \ref{fig:schematic_LM}. 

\begin{figure}[!h]
\centering
\includegraphics[width=0.95\linewidth]{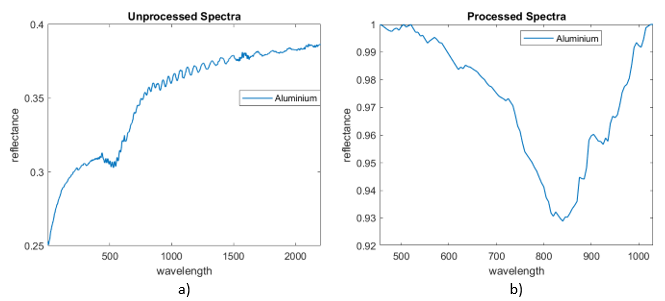}
\caption{Example of pre-processing where a) original Aluminium spectra and b) is spectra after all pre-processing steps}
\label{fig:Al_processing}
\end{figure}

As with the machine learning models it is necessary to first remove any atmospheric and solar illumination effects to achieve good performance although it is assumed this occurs prior to implementing the method in Figure \ref{fig:schematic_LM}. Band selection is used to select and use only wavelength bands for which information exists in the received spectra and all spectra in the library. In addition, the non-machine learning methods deploy data smoothing and continuum removal techniques to ensure that unmixing is based on key features. A detailed description of these techniques, with illustrating figures, can be found in \cite{vasile2022intelligent} 
Data smoothing will remove high frequency noise while continuum removal, achieved using a convex hull transform, will remove the overall concave shape often present in spectra while normalising the signal. This must be applied to both the library spectra and the received signal. This was not necessary in the machine learning technique,  as the network learns which features are important during training. An example of spectra before and after this processing can be seen in Figure \ref{fig:Al_processing}. Note that the overall signal has been flattened, high frequency noise removed and thus the key absorbance band at ~550nm is more pronounced.

\begin{figure}[!h]
\centering
\includegraphics[width=0.65\linewidth]{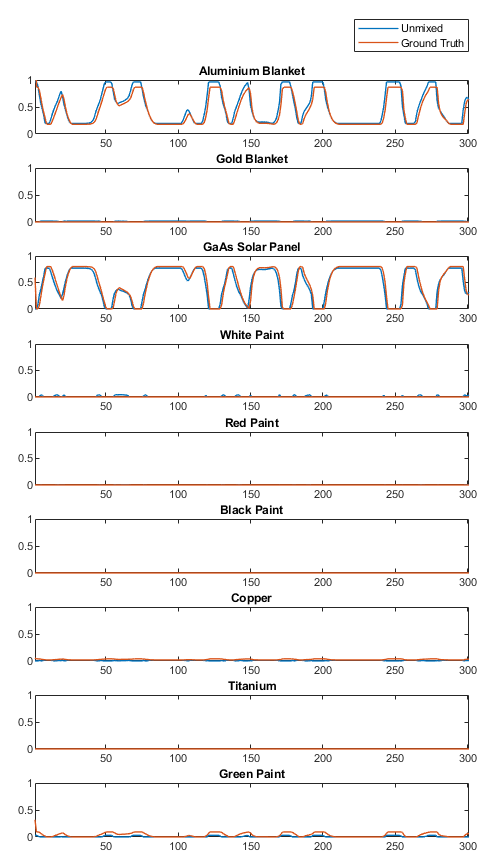}
\caption{Example of unmixing a received timeseries signal}
\label{fig:Unmixing_Regular_Example}
\end{figure}

The unmixing of the spectra based on the library can theoretically be performed using multiple unmixing techniques. In this work, methods based on non-negative constrained least squares (NCLS) and Joint-Sparsity and Total Variation (JSTV) \cite{aggarwal2016hyperspectral} were evaluated, and results were found to be very similar. The unmixing approach was validated on a small dataset of simulated data. An example unmixing result based on NCLS is shown in Figure \ref{fig:Unmixing_Regular_Example}.

The results show good accuracy with minimal or no estimated contribution from materials that are not present in the data. The library unmixing is also able to track changes in the abundance of materials that are present as the satellite rotates well, giving results that appear similar to those in the machine learning method. A full comparison between these methods is provided in Section 4.3.
To decompose the spectra into abundances of materials, the library mixing must find the optimal weighted combinations of materials to reconstruct the received signal and thus computational complexity scales with the number of materials in the library. In the case of the ANN method these combinations are checked in advance during training so the computational cost comes at that stage rather than during deployment, when the model most simply map the received signal to closest learnt combination. For the least squares based library unmatched the increase in complexity from larger libraries can significantly change execution time of the software. For example the 9 material library used in this study would take approximately 3 ms to unmix each spectra, while increasing the library 8-fold to 72 materials would increase computation time 50-fold to 140 ms.  
\subsection{Unmixing with Incomplete Spectral Libraries or Unknown Materials}

A limitation of library unmixing methods is the requirement to have an accurate spectrum in the library for each material that may be present or a large enough library to contain all possible variations of the spectra for each expected material. Given that the scope of the unmixing is to identify materials typically used in the space sector, building a sufficiently rich library with all representative materials is possible.

However, if a space object contained materials that were not present in the library it would not be possible to identify those materials. Furthermore, as the objective of the unmixing algorithm is to recreate the received signal as accurately as possible using weighted linear combinations for the library spectra, the algorithm would likely adjust contributions of other materials in an attempt to recreate the features of the missing ones. It is also possible that the aging of materials in space could result in differences in the spectra that are included in the library. In this section a preliminary exploration of both challenges is presented alongside a potential solution through the analysis of a residual portions of the signal that cannot be reconstructed.

\begin{figure}[!h]
\centering
\includegraphics[width=0.95\linewidth]{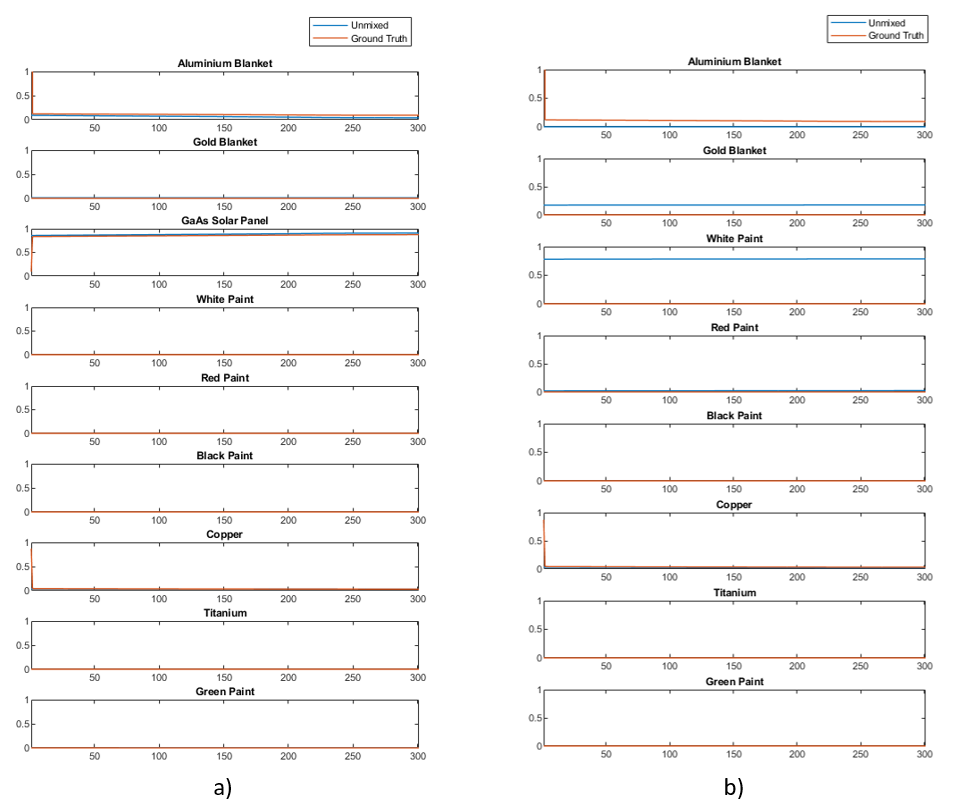}
\caption{Example of unmixing a received timeseries signal of Starlink where a) is unmixed with a complete spectral library and b) is unmixed with GaAs solar panel removed}
\label{fig:Starlink_Panel_Ex}
\end{figure}

The effect of materials that were not included in the spectral library can be studied by removing spectra from the library and using the algorithm described above to unmix the data described in Section 3. \ref{fig:Starlink_Panel_Ex} shows the result of unmixing a Starlink satellite with and without the solar panel in the spectral library. It can be observed that with solar panel present the abundance estimation is good but that when the solar panel is not present in the library the algorithm attempts to compensate for this by mixing in a combination of white and black paint.

\begin{figure}[!h]
\centering
\includegraphics[width=0.55\linewidth]{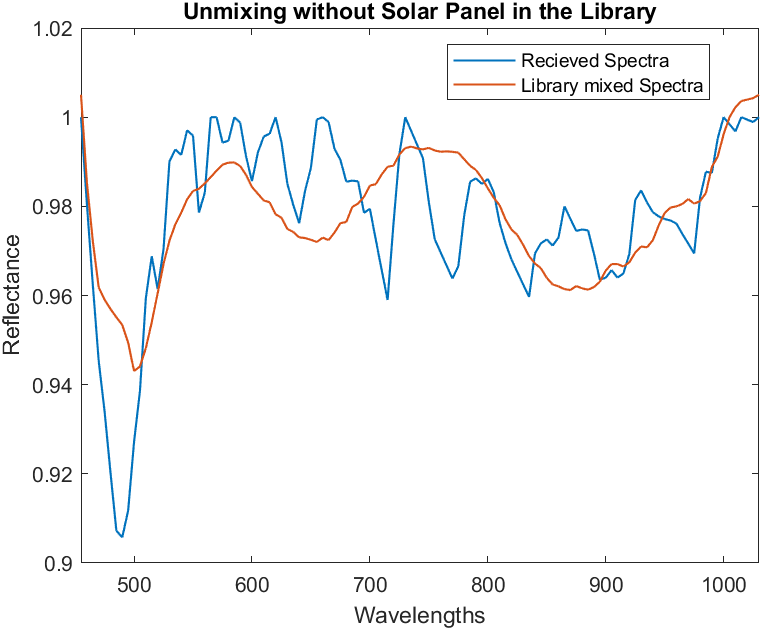}
\caption{Accuracy of unmixing using white and black paint to compensate for missing solar panel}
\label{fig:mixedSig}
\end{figure}

The reason for this can be explained by studying the received spectra after processing (blue) in Figure \ref{fig:mixedSig}. The received spectra has features which the unmixing model recreates as accurately as possible by using combinations of spectra in the library. Combining white and black paint allows the model to recreate the overall shape of the spectra, but misses fine features that are unique to the solar panel. 

\begin{figure}[!h]
\centering
\includegraphics[width=0.55\linewidth]{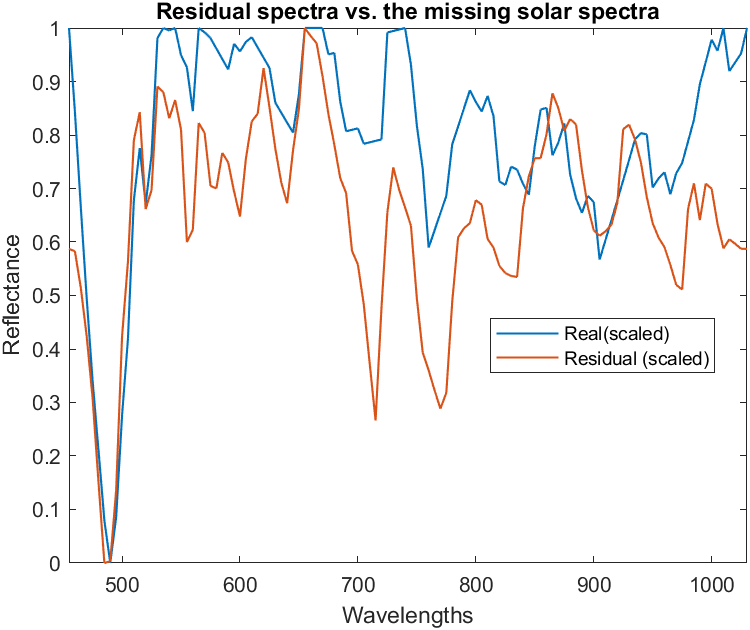}
\caption{Analysis of residual spectra compared to real solar panel spectra that was missing (scaled to show similarity in features)}
\label{fig:ResVsSolar}
\end{figure}

This behaviour is problematic and highlights the importance of having as complete a spectral library as possible when using this technique. However, this result also highlights potential solutions to mitigate this issue. Due to the missing material the mixed spectra did not match the input spectra well, thus the mean squared error between the received and mixed signal could indicate some cases were errors occur and that the result should be studied in more detail. As key features of the received spectra were not recreated in the mixed spectra the residual spectra (i.e. the difference between the mixed and received spectra) would capture these features. Features in the residual spectra will be caused by noise, errors in unmixing and missing spectra from the library. When features result from the presence of a material that is not in the library it would be expected that the residual spectra would resemble the spectra of the missing or unknown material. This concept is demonstrated in Figure \ref{fig:ResVsSolar}, which shows a comparison between the residual spectra in our Starlink example and the spectra of the solar panel known to be in the received spectra but not in the library. Note that as residual spectra are small and can have positive or negative values the spectra have been scaled to allow comparison of the features present. It is clear in this example that there is significant overlap between the residual features and the solar panel spectra, although these are not identical, likely due to the presence of noise and small unmixing errors.

\begin{figure}[!h]
\centering
\includegraphics[width=0.65\linewidth]{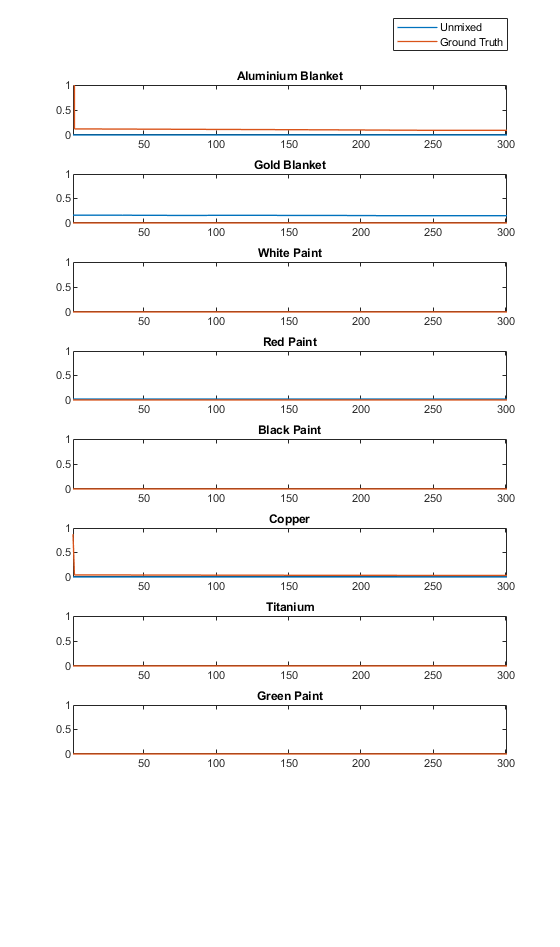}
\caption{Library unmixing of Starlink satilite with residual correction}
\label{fig:Starlink_noSolar_residualCorrected}
\end{figure}

It is important to note that in such cases the spectral library is incomplete and the primary use of this residual spectra should be to facilitate further analysis, understand what spectra might be missing and update the model. However, there is potential to use the residual spectra as a short term fix for the overfitting of white and black paint in this case. To do this the unmixing model is applied twice. On the first pass unmixing is based only on the spectra library and the residual spectra computed. On the second pass the residual spectra is normalised and included in the spectral library, thus allowing some abundance to be assigned to this rather than a material that is not present. The result of applying this residual based correction method to the Starlink unmixing is shown in Figure \ref{fig:Starlink_noSolar_residualCorrected}.

These preliminary results suggest good performance in preventing this overfitting problem by using the residual to correct the initial unmixing. However, the performance of this method may depend significantly on which signal was missing, or indeed how many signals were missing. In this case, the missing solar panel was a very significant material and thus the residual spectra matched its feature well. If instead the missing material made up only 1-2\% of object then this method may not help identify this material as well, as even 1\% of the original spectra resulting from noise, or a 1\% error in mixing, would create a residual that depends as much on these random processes as the missing material. 

\subsection{Comparison Between Machine Learning and Non-machine Learning Unmixing}
\begin{figure}[!h]
\centering
\includegraphics[width=0.85\linewidth]{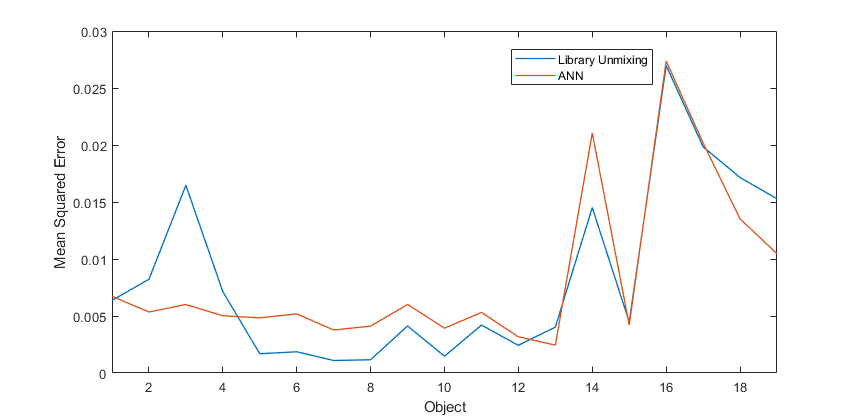}
\caption{Mean squared error of ANN and library matching unmixing approaches for different simulated objects)}
\label{fig:MSEvs}
\end{figure}

\begin{figure}[!h]
\centering
\includegraphics[width=0.95\linewidth]{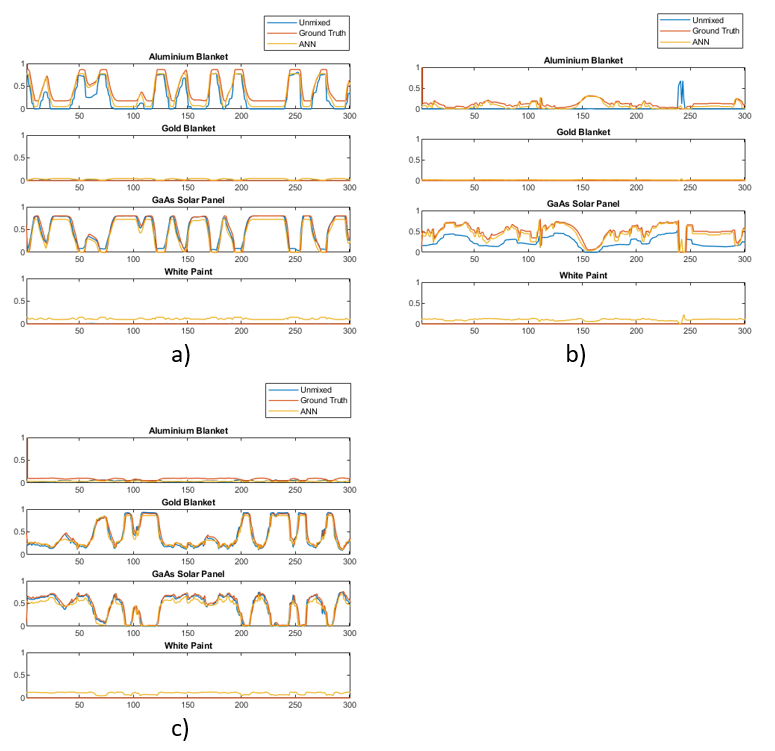}
\caption{Unmixing performance for aluminium, gold and solar panel for 3 objects with ANN and library matching methods where a) similar performance, b) ANN slightly better and c) ANN slightly worse}
\label{fig:compComb}
\end{figure}

To compare and analyse the differences between ANN and Non-machine learning library matching techniques a new dataset was generated with a few examples of various object types. Results from both methods on a larger dataset is also shown in Section \ref{sec:classification}. Both methods were used to unmix the spectra in the timeseries and the resulting abundance estimations were compared to the known ground truth. Results are broadly similar which is confirmed by the similarly low mean squared errors shown in Figure \ref{fig:MSEvs} and the similar abundance graphs in  Figure \ref{fig:compComb}, which shows examples of unmixing results for three different objects. The mean squared error of the ANN was 5.3 $\times$ 10$^{-3}$, nearly exactly the same as the previous trial, and the mean squared error of library unmixing predictions was 4.5 $\times$ 10$^{-3}$. This is a small difference and Figure \ref{fig:compComb} shows the the best unmixing result differs from object to object, suggesting either model could be used to obtain similar results and which method has the lowest error would vary between datasets. However, while the overall magnitude of errors are similar the type of error encountered by both methods are quite different. Figure \ref{fig:compComb} a) shows a typical example were differences between the methods are very subtle while b) shows the case where the ANN performed slightly better than the library matching and c) shows the case where the ANN performed slightly worse. It can be observed that when the ANN produces the errors it typically does so by predicting a small amount of a material that is not there (such as white paint), while the errors in library matching more often come from over- or under- estimating particular abundances, or missing a material that exists in a smaller quantity. This suggests library matching may produce larger errors, but less frequently compared to the ANN. Overall the performance of both methods is found to be accurate and reliable for further classification and analysis, however, these differences in characteristics are worth considering in future developments. For computational complexity, the non- machine learning library unmixing takes around 3 ms to analyse each spectra compared to 0.7 ms for the ANN method. This performance advantage for the ANN is likely due to more of computation taking place at the training stage of the ANN, thus this advantage will increase further as the complexity of the problem increases, such as if larger material libraries were used.

\subsection{Material Aging and Inaccurate Library Spectra}

Materials in space exhibit spectral changes over time due to exposure to the space environment \cite{duzellier2022space, pearce2020examining}. Although the ANN could easily be retrained with a new set of spectra corresponding to a given exposure duration, if the object being imaged is not already well-known, it is unlikely that good knowledge of its age will be available. Additionally, to the best of the authors' knowledge, no models currently exist to describe the spectral aging process and predict the spectral changes that will occur over time. Thus there is likely to be some error in the library spectra with respect to the underlying materials on the real object. It is, therefore, important to understand the effects of such error.

\begin{figure}
    \centering
    \includegraphics[width=0.65\textwidth]{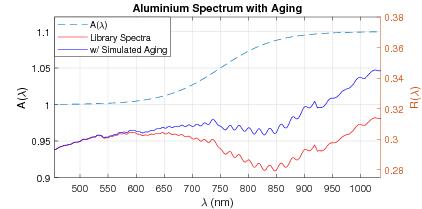}
    \caption{Aging function (with $a = 0.1$) and its application to the spectrum of aluminium blanket, in the wavelength region of interest for this paper (455 - 1035 nm)}
    \label{fig:aging_example}
\end{figure}

 Since no model is available, as a first-order approximation of the spectral changes we define an aging function $A(\lambda)$, which is used to transform the reflectance spectra $R(\lambda)$ of all materials in the library: $R(\lambda) \rightarrow R(\lambda)A(\lambda)$. The function was chosen to accentuate reflectance in the red or near infrared spectral regions relative to shorter wavelengths given that in the literature there is the evidence that the reflectance spectra of aged materials display a shift to the red. 
 An aging function of sigmoid shape is defined such that an increasing boost is given to the spectrum in the red and infrared regions:

\begin{equation}
    A(\lambda) = 1 + \frac{a}{1+e^{-0.02(x-750)}}
\end{equation}
We reiterate that this is simply a first order approximation of the spectral reddening effect reported in some literature on object spectrometry and is not to be taken as an accurate physical model of the process. The aging function along with a comparison of aged and unaged aluminium blanket spectra are illustrated in Figure \ref{fig:aging_example}.

\begin{figure}
    \centering
    \includegraphics[width=0.95\textwidth]{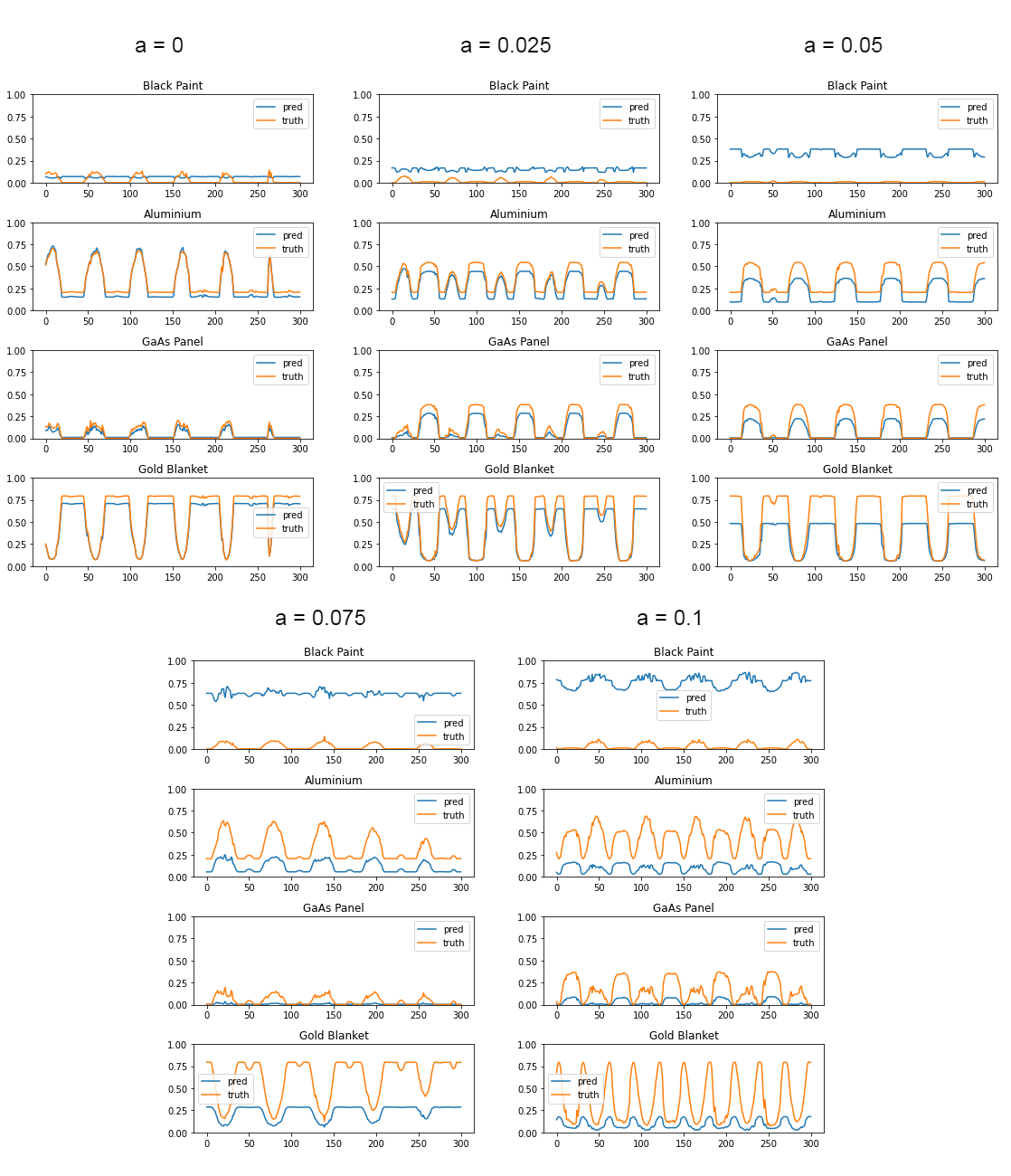}
    \caption{MAC model predictions compared with true values for DubaiSat-2 with progressively increasing spectral aging. Only present materials shown. Different simulations of the same satellite shown in each subfigure.}
    \label{fig:dubaisat_aging_study}
\end{figure}

To investigate the effect of this spectral error on the ANN decomposition, we generated several datasets of simulated measurements for DubaiSat-2, where the underlying materials in the simulation model have been replaced by increasingly aged versions ($a = 0, 0.025, 0.05, 0.075, 0.1$). The existing ANN (trained on unaged spectra) was then used to decompose the signal into its material components. Model predictions for these four aged versions of DubaiSat-2 are shown in Figure \ref{fig:dubaisat_aging_study}.

These results show that discrepancies between the spectra used to train the MAC model and the spectra of the materials in a real object do cause errors in the predictions made by the MAC model. Of particular note is the progressively increasing overestimation of the abundance of black paint, as the aging is increased. Since the sum of the material decomposition is constrained to 1, this causes other materials' contributions to be reduced to compensate. However, it should be noted that the overall shape of the decomposed material curves is preserved.

\begin{figure}[!h]
\centering
\includegraphics[width=0.95\linewidth]{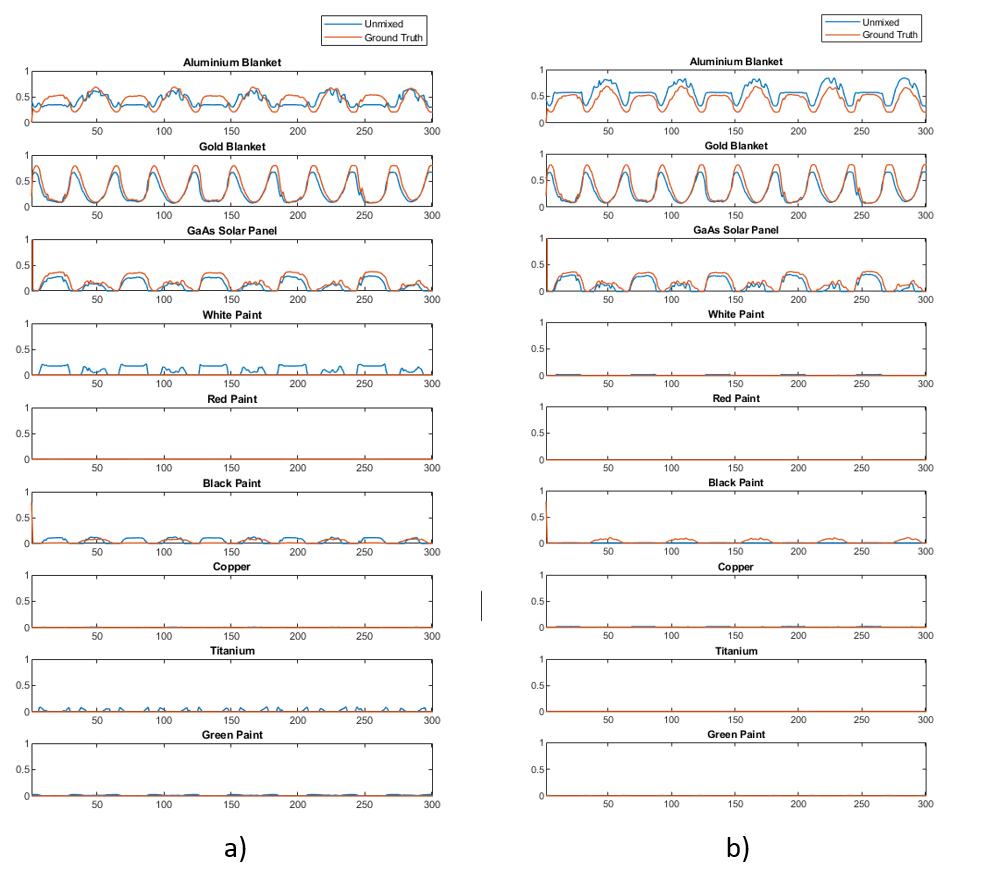}
\caption{Unmixing of Aged DubaiSat with a) the original spectral library, b) a spectra library that has also been aged}
\label{fig:libAged}
\end{figure}

Repeating this test with the library matching method show similar results with typical examples of the error shown in Figure \ref{fig:libAged} a). Here it is observed that the overall estimate of material abundance is accurate, but changes in the abundance at some time points are missed. This is particularly noticeable in the aluminium. The relative performance difference here is likely explained by the pre-processing used, which includes data smoothing and continuum removal, which will reduce some of aging effects. However, errors still exist and is likely caused by aging creating more similarities between the spectra than before. As with the ANN method it is possible to re-train with knowledge of the aged spectra to obtain better results as shown in Figure \ref{fig:libAged} b). In this case that is achieved by adding aged spectra to the library in place of the original spectra.

\section{Probabilistic Classification of Space Objects Based on Material Identification}
\label{sec:classification}

Once a model is developed to extract the material abundance curves (MACs), its output can be further processed to extract information on the underlying space object. Specifically, we propose that knowing the abundance of material on the surface of a space object can allow the class of that object to be predicted, which in some cases may inform the purpose of the object itself. In theory, space objects with a similar general purpose are likely to have similar visible components, and thus similar materials. For example, satellites are expected to have solar panels while rocket bodies would not and thus the presence of any material from which solar panels might be made will help predict which of these we may observe. Similarly, those satellites deployed for communication purposes would be expected to have antennae, so materials from which an antenna might be made, could be used to help predict which satellites fit this classification. It is desirable to design a classification system that is agnostic to attitude and pointing direction, as the resulting model would then be applicable to multiple deployment scenarios. Thus the model used to simulate satellites will randomly generate its motion and not assume the same repeated behaviour. While the abundance of materials will very likely help distinguish objects, the former assumptions means that it cannot be guaranteed that the surfaces of the satellite that are observed will accurately reflect the true abundance. Indeed in any practical scenario this is also unlikely to be the case. For this reason, the classification model is designed to classify objects based on the probability of detecting materials. 

To achieve this, we first demonstrate that the presence of materials on space objects can be reliable identified using the MAC developed in previous sections. A probabilistic approach is used in which the time-series data of material abundance is mapped to a single probability score indicating the confidence that the material is present on the object, rather than a small false positive abundance due to noise. This probability estimate for each material is then used to classify the space object. While it would be possible to train on known object data from simulations, this has the potential to create over-fitting problems and may lead to a model that does not generalise well. Instead the proposed classification model is trained on synthetic instances representing the potential material probabilities of each class. It is also possible that a machine learning model trained on simulated objects might learn to identify each object and then map this to the corresponding task, which is the opposite of the intended workflow where we intent to perform a more general classification to progressively reconstruct a full picture of the observed object. Thus the model we train has never seen any simulated or real object data until the moment it is tested, as all training data are only synthetic mixtures of materials.

To be noted that the composition of the space objects used in the simulations in this section does not necessarily correspond to the actual composition of those objects. Shapes, attitudes and material distributions were chosen to be sufficiently representative of existing space objects but do not correspond to the truth.  It is, therefore, possible that the objects listed in this section are not really covered in the materials we selected nor the abundance of those materials corresponds to the actual one. On the other hand the examples in this section are useful to demonstrate the ability of the proposed methodology to identify the materials and classify the corresponding object because we can compare the outcome of the proposed pipeline against the ground truth of the objects we simulated.

\subsection{Probabilistic Material Classification}
The first stage in this approach is to analyse the material abundance estimation, to gain a probabilistic estimate of what materials are present. The material abundance models proposed in this work do not inherently provide a binary material presence decision. This is because these are unmixing models that attempt to find the proportions of each material from a known library that in a linear combination would produce the received spectra. Thus the probability of a material being present is linked to the abundance of the material that is estimated and must be extracted from this for our purposes. As it is known that the visibility of any material will alter as the object rotates the maximum abundance of each material detected for any point in time offers the best estimate of being able to conclude if that material is present. The more significant the features of a given material spectra are in the received spectra the more confident the model is that this material is present and subsequently a greater abundance is predicted. However, it is not helpful to consider 100\% abundance as the threshold for 100\% probability as most materials on most objects never reach this abundance for any angle of observation. Instead a far lower abundance should be sufficient to provide confidence that a material is present. In the ideal case the abundance that equates to 100\% probability of prediction should be the lowest abundance that could not be generated by noise. We propose a threshold where the model has maximum confidence that a material exists and linearly scale the output of the material abundance prediction model accordingly. This allows separate measurements of probability of detection and material abundance, despite being derived from the same data. For example, with the threshold set to 5\% and two satellites had 10\% and 50\% solar panel respectively, the model would return a 100\% probability of the panel existing in both cases, but retain the knowledge that there is a larger panel in the second case. Empirical studies of the error rate of different values suggest a threshold of 5\% appears effective in this study as it is consistent with the smallest abundances observed and greater than the typical error rate from the material abundance models. This would mean that the 1-2\% false positive abundance predictions discussed previously would equate to a relatively low probability of predicting that material is present.

\begin{figure}[!h]
\centering
\includegraphics[width=0.95\linewidth]{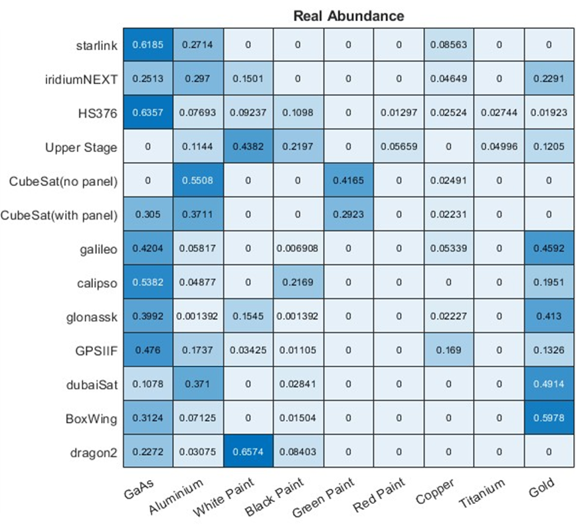}
\caption{Ground truth mean abundance across 2000 samples of each satellite of a material being present on that satellite}
\label{fig:gtMatAbund}
\end{figure}

To investigate these probability estimates, a large dataset of 26,000 space objects were generated, 13 object types with 2000 simulated instances of each. From this simulation we record the abundance of each material that is visible to the observed and used by the model to generate the spectra. The mean abundance of each material for each type of object is shown in Figure \ref{fig:gtMatAbund}. All objects were observed for 5 minutes at 1 Hz with the simulation using the same observation and rotational conditions described in \ref{sec:simulations} and \ref{tab:OE_randomized}.

\begin{figure}[h]
\centering
\includegraphics[width=0.95\linewidth]{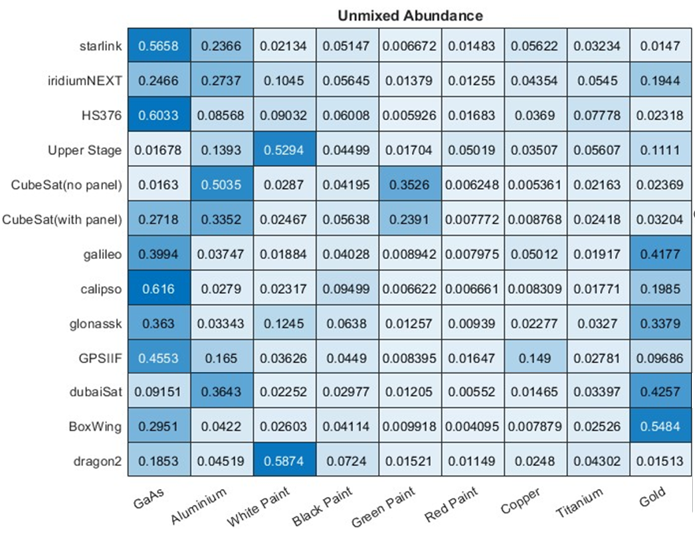}
\caption{Mean abundance of materials across 2000 samples of each satellite of a material being present on that satellite according to the ANN model}
\label{fig:Mat_abund}
\end{figure}

\begin{figure}[h]
\centering
\includegraphics[width=0.95\linewidth]{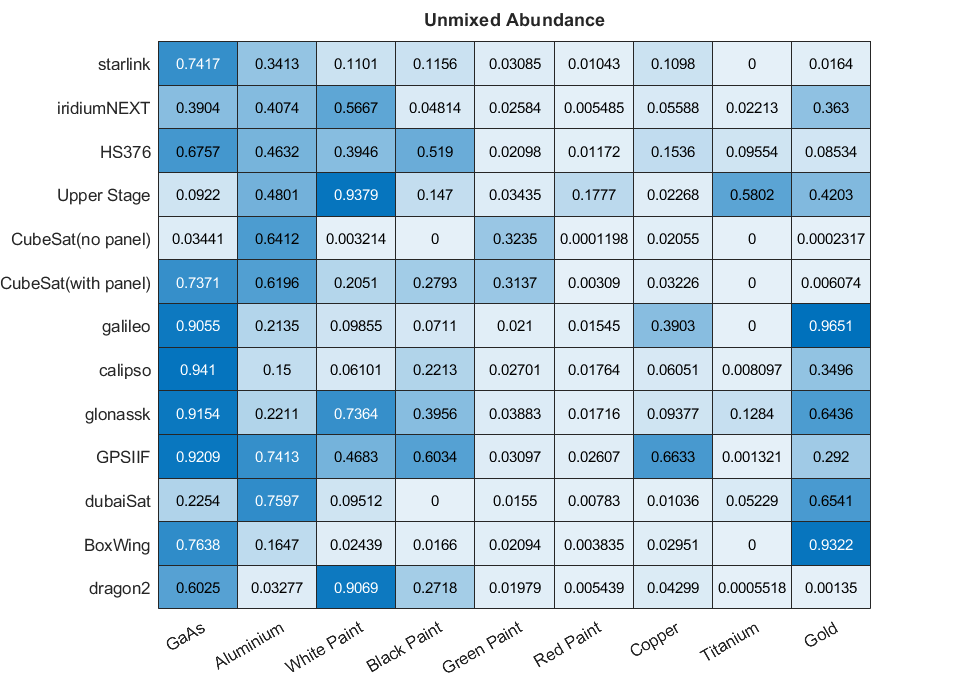}
\caption{Mean abundance of materials across 2000 samples of each satellite of a material being present on that satellite according to the library unmixing model}
\label{fig:Mat_abund_lib}
\end{figure}

Using the MACs produced from the same simulated data using the proposed unmixing model the predicted mean abundance of each material in each satellite can also be computed using either the ANN or library matching unmixing models, and is shown in Figure \ref{fig:Mat_abund} and Figure \ref{fig:Mat_abund_lib}, respectively. 

These results show particularly good agreement between the ground truth and proposed ANN model as to which materials are most abundant in each object. While both method unmix the same general trend the library unmixing methods general tend to over-estimate the maximum abundance observed for each material. However, it is rare for the unmixing models to predict 0\% abundance of any material, instead often finding a relatively low abundance of the materials that are, in reality, not there. For example, most objects in the dataset contain GaAs due to the solar panels present on these satellites. Both models predict this well including correctly identifying relative abundances, such that Starlink satellites have around twice the amount of this material as Iridium-NEXT satellites. However, the two objects without GaAs have a prediction of 1.6\% GaAs abundance with the ANN model, and higher in the library unmixing. This is higher than the true abundance of 0\% but far lower than any object that really did have this material. Similar trends can be observed throughout this experiment. 

\begin{figure}[!h]
\centering
\includegraphics[width=0.95\linewidth]{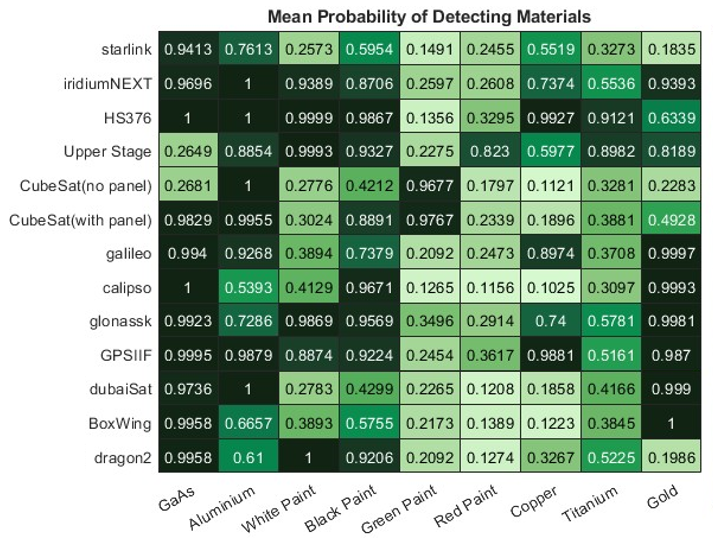}
\caption{Mean probability across 2000 samples of each satellite of a material being present on that satellite according to the ANN model}
\label{fig:Mat_probs}
\end{figure}

For classification of satellites the critical factor will be how well our material probability estimates can distinguish false from true abundances of material, particularly for materials such as copper, where the true abundance to be detection is typically very low. As the ANN model achieves the most consistent performance, this model is used to estimate the probability of detection of each material. The probability of detection computed by the proposed model is shown in Figure \ref{fig:Mat_probs}. Note that a probability score indicating how likely each material is to be present is compute for each instance of each object, with the mean score across the 2000 instances reported here. Thus 25\% does not mean that the material is predicted 25\% of the time but that on average that object has a 25\% probability score of that material being present, which if consistent may result in the model never predicting that the material is truly there.

The material probabilities correlate well prior expectations. Upper stages which are known not to contain solar panels, show a low probability of GaAs, as does the CubeSat that we not to contain panels. All satellites known to contain solar panels on the other hand show high probabilities of GaAs. So while there was some false positive abundance the probabilities clearly distinguish those objects that have this material from those that do not. Materials such as copper, that always exist in lower abundances in our dataset have greater uncertainty, with more objects showing a mean probability of prediction closer to 50\%. However in the majority of cases the probability scores correlate well with the objects that do have copper, with the exception of upper stages were false positive copper defections are relatively likely. In other cases the probabilities demonstrates were uncertainty is likely to occur. Similarly, in the test dataset used here titanium exists only on HS376 and the Upper Stage. It is evident here that the probability of detection is significantly higher in these two examples, but there is also a reasonable probability of detection in a few other satellites, suggesting there are features there that might mimic titanium. 

While it is useful to study the probability of materials present it is also useful to retain information of the abundance of each material, as shown in Figure \ref{fig:Mat_abund}. For example, when comparing this information it can be observed that while a similarly high probability of GaAs is found in Starlink and Dubaisat the model is also able to predict that Starlink has on average twice as much of that GaAs material as the DubaiSat. Abundance is estimated at every point in time though and in this instance it is the maximum abundance that is shown. It may be useful to study the mean abundance to assess the overall composition of the satellite, or to attempt to study particular faces of a satellite. For example, the time where GaAs abundance is maximised should indicate the solar panels facing towards the sensor. It may then be the case that the it is known what materials should be visible from that perspective to help the classification. 

\FloatBarrier

\subsection{Probabilistic Object Classification}
The final stage in this work is to classify satellites into classes that describe their general composition or intended purpose. We use only the probability of each material being present on the object to minimise the dependency of the model on the satellites movement or the visibility of certain faces of the satellite. The previous section suggests that the ANN method produced more accurate results, thus the object classification methods uses the probability of detection from only the ANN method as in inputs.  While the abundance of material present is theoretically also a useful predictor, this is intentionally not used in the current model. This is because the abundance of material that is measured correlates with the randomised movement of the simulated objects, i.e. measuring half the expected abundance may simply mean the component made from that material is partially obscured during observation. Therefore, in this initial classification model the abundance is not used and instead it is explored how accurate a classification could be made using only the previously derived probability of each material being present.
For this trial satellites are split into the classes Comms/GNSS, Rocket Body, Cubesats, Earth Observation (EO) and Capsule as defined in Table \ref{tab:satellite_class_list}. Note that the Boxwing satellite is a generic shape so has no obvious class.

\begin{table*}[!t]
    \centering
    \begin{tabular}{c|c}
        Satellite & Classification\\ \hline
        Starlink & Comms/GNSS\\
        Iridium-NEXT & Comms/GNSS \\
        HS376 & Comms/GNSS \\
        Galileo & Comms/GNSS\\
        GPS Block IIF (GPSIIF) & Comms/GNSS\\
        GLONASS-K & Comms/GNSS\\
        Upper Stage & Rocket Body\\
        DubaiSat-2 & EO \\
        CALIPSO & EO\\
        CubeSat (panels) & CubeSat\\
        CubeSat (no panels) & CubeSat\\
        Box-Wing & None\\
        Dragon2 & Capsule\\
    \end{tabular}
    \caption{Full list of satellite models used and corresponding classes}
    \label{tab:satellite_class_list}
\end{table*}

As mentioned previously training a model on examples of these objects achieves high classification accuracy but also leads to over-fitting the model to the satellites in the training set. To avoid this the model is instead trained on synthetic combinations of material probabilities. This has the advantage of not using satellite data during training and allows the full range of combinations of material probabilities to be used, not just those combinations that occur in the training datasets. 

\begin{figure}[!h]
\centering
\includegraphics[width=0.95\linewidth]{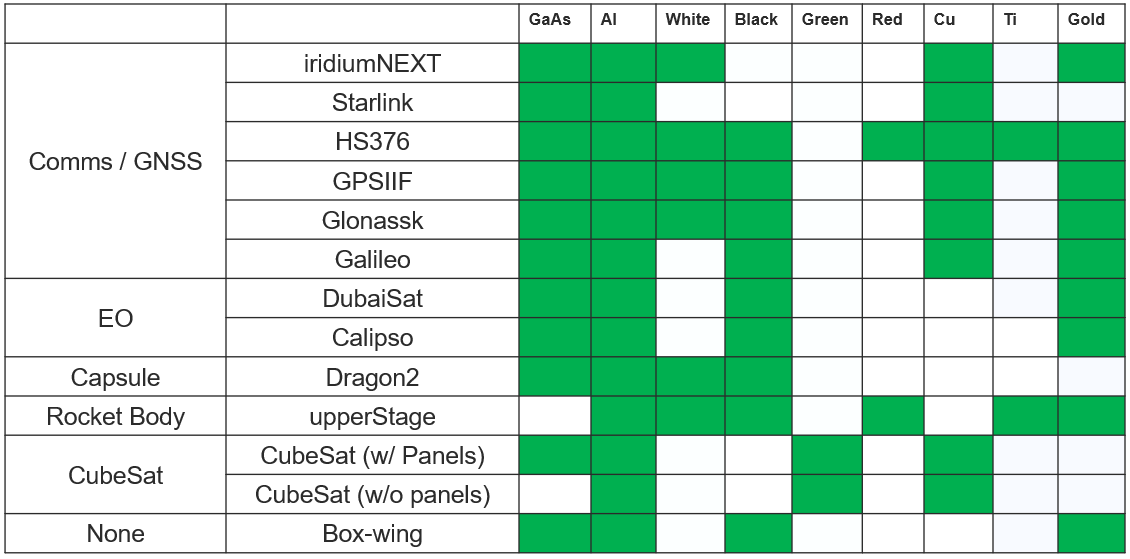}
\caption{Materials that are expected to be present on each object in the simulated dataset}
\label{fig:satMatTruthTable}
\end{figure}

Synthetic data is generated by first defining a probability range for each material in each class, that specifies the highest and lowest probability that may occur in measurements. To do this it is necessary to make assumptions about the materials in each class. This is achieved by studying the materials that are expected to be present on each satellite, as shown in Figure \ref{fig:satMatTruthTable}. 

For example, if it is assumed titanium is present in all rocket bodies then a suitable range for synthetic data for rocket bodies might include titanium probabilities in the range 50-100\%. Meanwhile, for Comms/GNSS satellites Titanium is present in some but not all objects, so the synthetic data for the Comms/GNSS class may include titanium probabilities in the range 0-100\%. When generating this data it is helpful to ensure all extrema of these ranges are included in the synthetic data. The training data used in this study is of the ranges shown in Table \ref{tab:synthDataTable} and for each class every combination of material probabilities is generated with the limit that only the maximum, minimum and median value of the defined range are used. This resulted in 358,000 synthetic combinations being used in the training. Optimising this synthetic training data based on effect on the subsequent object classification model would risk the over-fitting scenario this approach intended to avoid. However, some observations about the original material detection model were used to define optimal ranges for the table in Table \ref{tab:synthDataTable}. It can be observed from Figure \ref{fig:Mat_probs}, that false positive probabilities >40\% are likely to be rare in the proposed model. So when generating synthetic data the probability generated for non-existent material is defined so as to not exceed 30\%, while when materials are present values between 50-100\% are typically generated. For copper, which is harder to predict given the low abundance, the generated training data includes probabilities of prediction as low as 40\%, when copper is present, to reflect this low abundance. Similarly for solar panels the generated data includes only probabilities above 90\%. These threshold help generate training data but are not threshold on the model, which will learn and apply its own classification thresholds based on the relationship between these datasets.

This is an arbitrary definition of the classes but it is useful to group objects based on their composition. The scope of this definition of the classes in this paper is to show an example of how to classify from the probability identification of materials and to illustrate the possible critical aspects of the classification process. Alternative definition are indeed possible.

\begin{table*}[!t]
    \centering
    \begin{tabular}{c|c|c|c|c|c|c|c|c|c}
        Class       & GaAs       & Aluminium  & White     & Black     & Green   & Red         & Copper     & Titanium  & Gold    \\ \hline
         Comms/GNSS  & 90-100\%   & 50-100\%   & 0-100\%   & 0-100\%   & 0-30\%  & 0-100\%     & 30-100\%  & 0-100\%    & 0-100\% \\
         EO          & 90-100\%   & 50-100\%   & 0-30\%    & 50-100\%  & 0-30\%  & 0-30\%      & 0-30\%    & 0-30\%     & 50-100\%\\
         Capsule     & 90-100\%   & 50-100\%   & 50-100\%  & 50-100\%  & 0-30\%  & 0-30\%      & 0-30\%    & 0-30\%     & 0-30\%  \\
         Rocket Body & 0-30\%     & 50-100\%   & 50-100\%  & 50-100\%  & 0-30\%  & 50-100\%    & 0-30\%    & 50-100\%   & 50-100\%\\
         CubeSat     & 0-100\%    & 50-100\%   & 0-30\%    & 0-50\%    & 50-100\%& 0-30\%      & 30-100\%  & 0-30\%     & 0-30\%  \\     
    \end{tabular}
    \caption{List of classes and range of material probabilities used in synthetic training dataset}
    \label{tab:synthDataTable}
\end{table*}

Comms/GNSS satellites and EO satellites are clearly similar but should be distinguishable by the copper components. Similarly the paints present should distinguish Capsule and CubeSats from other classes, while the lack of GaAs should help identify rocket bodies. While the generic BoxWing satellite has no obvious classification it is most similar to those in the EO class so it would be expected that this would be the most probable classification.

\begin{figure}[!h]
\centering
\includegraphics[width=0.95\linewidth]{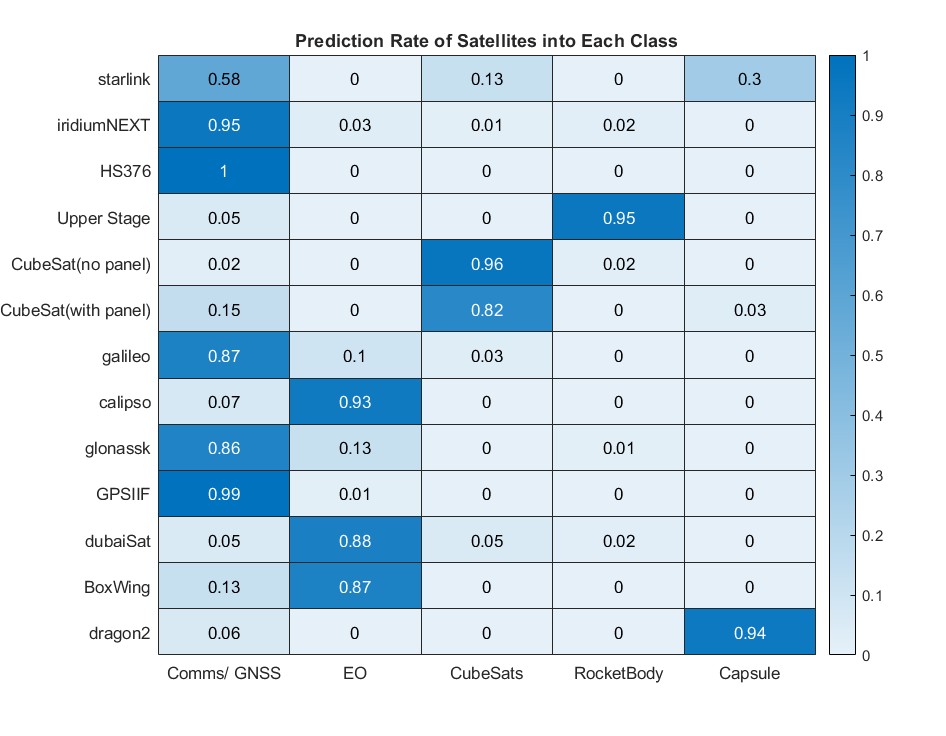}
\caption{Prediction rate from the knn model when classifying objects based on probability of materials present on the object}
\label{fig:classPredDe}
\end{figure}

A k-nearest neighbours (KNN) model is trained using the synthetic combinations of material probabilities defined in Table \ref{tab:synthDataTable}. The benefits of this method in this case is that classification is based on a distance in feature space between the observation and training data instances of each class. It is, therefore, well suited to classifying objects where the probabilities do not exactly match the ideal cases shown in the dataset. The prediction rate for the 2000 simulated instances of each satellites into the each class is shown in Figure \ref{fig:classPredDe}. 

In the majority of cases this classification model produces convincing results. Most classes are classified well and the overall classification accuracy across all satellites is 89\%. Across 12/13 objects classification accuracy is 92\%, as most errors are in Starlink satellites. As significantly, the classification errors are typically logical and explainable.  For example, confusion between Comms/GNSS and EO is the most common error, which is to be expected given the similarity of these objects and that only small abundances of copper typically distinguish them. Conversely, it is quite rare for satellites to be misclassified as rocket bodies, which is expected given the absence of solar panels, which are typically quite significant features. Boxwing, which is a generic object with no obvious or defined class class, is predicted as EO 87\% of the time, as was expected given the similarity in materials. 

In the Starlink class, accuracy is 28\% lower than the next least accurate class. A significant number of Starlink objects are predicted as Capsules in particular, one reason for which may be a failure to detect the copper in many cases. Copper is a challenging material to detect as it often occurs in low abundances and the decomposition model returned lower probability of copper presence then any other object known to contain it, as shown in Figure \ref{fig:Mat_probs}. However, when investigating the ground truth  data of the observation of spectra from simulated test objects, it is found that in 37\% of Starlink objects no copper was present. This occurred as the component made from copper was completely obscured in these cases and did not face the observer. Thus the errors here are explainable, and while it is probable that some errors results from visible copper that is not properly detected the most significant error is that the classification is likely based on a material that was not visible in all starlink observations, so could not possibly be detected. This highlights a potential limitation of classifying satellites based only on the material present when those materials are not guaranteed to always be visible. However, the same problem could theoretically occur in other objects where the error rates were far lower, so it is reasonable to assume the geometry and materials present in Starlink are more challenging to classify.

\begin{figure}[!h]
\centering
\includegraphics[width=0.95\linewidth]{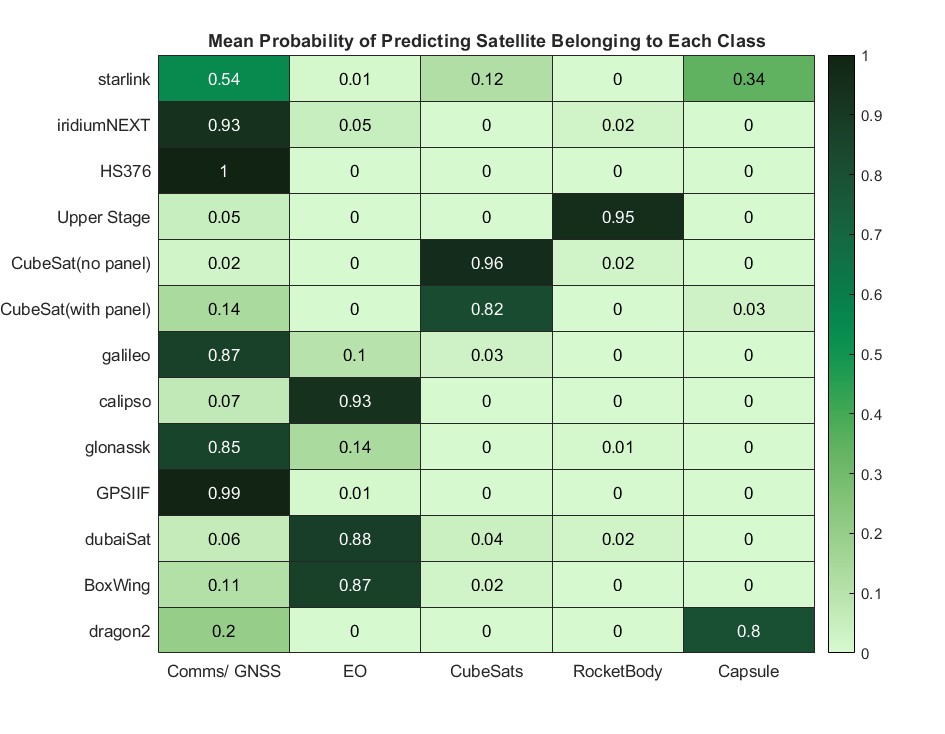}
\caption{Mean probability of predicting object as belonging to each class in the KNN model}
\label{fig:classPredProb}
\end{figure}

Similarly to earlier in the pipeline, where material presence was estimated, the probability of each classification is also interesting to study and provides more explainability to the model. The KNN algorithm classifies each combination of materials based on the distance between the observed object and all the classes identified through the training data. This means that a relative metric of the probability of the observed object belonging to each class inherently exists within the model. The average of these classification probabilities across the 2000 instances of each object type are shown in Figure \ref{fig:classPredProb}. 

The mean probabilities are quite similar to the actual error rate, which implies most predictions are made with high confidence. However, the usefulness of this metric alongside classification results can still be observed. For example, this reveals that while 94\% of Dragon2 objects were correctly identified as capsules, on average the confidence of this prediction was 80\%, which is lower than the confidence of many other predictions, with a mean probability of 20\% that this is a Comms/GNSS. This makes sense as the materials using in capsules are broadly similar with only a single material difference between Dragon2 and some Comms/GNSS satellites. It is encouraging that while this does not cause a 20\% error rate, the similarity between these classes and potential for confusion is still captured here.

\begin{figure}[!h]
\centering
\includegraphics[width=0.95\linewidth]{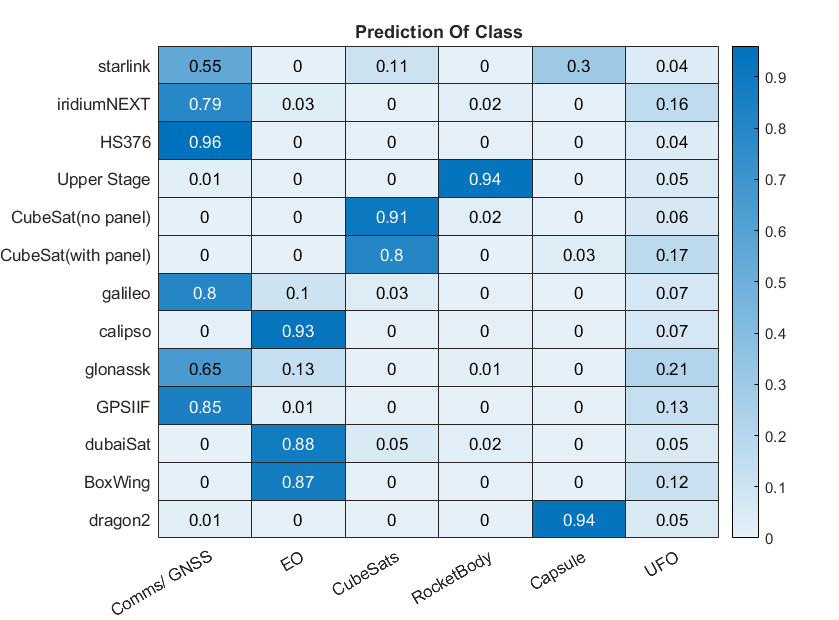}
\caption{Prediction rate from the knn model when using model confidence and prior assumptions to flag uncertain predictions as UFO}
\label{fig:classPredUFO}
\end{figure}

Another option for utilising probability data is to identify objects for which the correct class is highly uncertain and label these as a Unidentified Flying Object (UFO), rather than as whichever class the model deemed only marginally more probable. This is also very useful when unseen combinations of materials are present or materials not int he library are detected. Simple, manually derived checks may also be introduced here to ensure that each prediction matches key assumptions for each class. To demonstrate this theory we test the assumption that, a) solar panels (GaAs) are present in all communication satellites and never in rocket bodies and b) green paint is present only on Cubesats and not on anything else, are used as a final confidence check on the classification. Any classification that does not match these assumptions are instead labelled as UFO. The revised classification results using this method are shown in Figure \ref{fig:classPredUFO}. It is clear that this approach most significantly effects the Comms/GNSS class. False positive classifications in this class are almost completely eliminated, however, this occurs at the expense of failing to classify a significant number of Iridium-NEXT and Glonaask satellites, which instead are labelled as UFO. The true positive rate for this class fell from 87\% to 77\% using this approach while the false positive rate fell from 7\% to 0.3\%. Overall accuracy is lower due to the drop in true positive rate, however, the confidence in the remain classifications increases. In the proposed KNN model 91\% of Comms/GNSS classifications were originally correct, but with the UFO class included this increased to 99.5\%. This is a desirable behaviour because the classification system because the UFO class is, effectively, the uncertain class. Hence the classifier moves to the uncertain class all instances for which a decision has low confidence.

This provides an alternative classification approach depending on whether it is more important to identify as many objects correctly as possible or if is more important to ensure those classified are correct, at the expense of failing to classify some at all. The probabalistic approach to classification has other practical benefits not explored in this study. For example, the validation trials in this work assume a single observation of each object and no other pieces of information on orbit or motion. In practise, it is likely that objects classed with lower probabilities can be re-examined on multiple occasions to increase the information available and improve the confidence of the classification result.

\section{Conclusion}

In this paper we presented a novel data processing pipeline designed to determine the materials present on unknown space objects, and classify accordingly. This was comprised of a model to decompose a given hyperspectral signal into the relative contribution from each present material, which was shown to be possible using an artificial neural network, or alternatively, more traditional techniques such as constrained least squares. The behaviour of these approaches was also investigated when applied in non-ideal scenarios; when there is discrepancy between the training set and the real materials (for example, with weathered satellites), and the case where the object contains a material that is not present in the library. It is also demonstrated that the material abundance curves can be used to estimate the probability of each material being present. A second model was then developed which estimates the probability of a space object belonging to predefined classes. Based on the output of this model satellites are labelled either as belonging to a known class or being unidentified and labelled as a UFO.  Generalisation to unseen object was prioritised in the design of these models, with both the decomposition model and classification model being constructed in an inherently geometry-agnostic manner. Both models are trained using synthetic data that mimics the received spectra and output of the decomposition model respectively. This means that no space object data, real or simulated, is used to train the model. Validation on 26,000 simulations split across 13 types of space object, including simulation of atmosphere and randomised attitude, demonstrates good performance in predicting materials in classes. In cases where the object is classified incorrectly, the proposed pipeline is valuable, as by first analysing the materials present the reason for each classification decision can be understood. This makes classification results very explainable. Typically errors are due to low abundances, that are harder to detect, or because the material exists on a small component of the satellite that is not visible at all during observation. Techniques are used for which a probability score can be derived from the data or inherently exists within the machine learning model used. These probabilities would indicate any uncertainty in prediction that should be considered. By using two models that feed into each other, uncertainty in classification could be traced back to uncertainty in the presence of a certain material. This further enhances the explainability of this model.

Note that the definition of the classes in this paper serves only the scope to show how to group objects from their material composition. Thus it does not imply the specific function of any of the objects in each of the classes nor it defines what a EO or a comm object is. Alternative definitions are possible without changing the overall classification approach.

A complete classification system would, however, require an extended library, with proper consideration for material degradation, and a more in depth analysis of the resulting computational complexity. On the other hand, we showed that for materials that appear in large abundance the classification system working with a small library is quite robust and can provide a reliable association of objects to classes. When uncertainty is high an a deeper analysis is required the UFO class is effective at capturing undecided objects.

\section*{Acknowledgements}

This project is supported by the ESA OSIP grant HyperClass. The authors would like to thank Emiliano Cordelli and Jan Siminski of the ESA debris office for their advice and guidance.

\clearpage
\bibliographystyle{jabbrv_unsrt}
\bibliography{sample} 

\begin{thebibliography}{10}
\urlstyle{rm}
\expandafter\ifx\csname url\endcsname\relax
  \def\url#1{\texttt{#1}}\fi
\expandafter\ifx\csname urlprefix\endcsname\relax\def\urlprefix{URL }\fi
\expandafter\ifx\csname doiprefix\endcsname\relax\def\doiprefix{DOI: }\fi
\providecommand{\bibinfo}[2]{#2}
\providecommand{\eprint}[2][]{\url{#2}}

\bibitem{JORGENSEN20041021}
\bibinfo{author}{Jorgensen, K.} \emph{et~al.}
\newblock \bibinfo{journal}{\bibinfo{title}{Physical properties of orbital
  debris from spectroscopic observations}}.
\newblock {\emph{\JournalTitle{Advances in Space Research}}}
  \textbf{\bibinfo{volume}{34}}, \bibinfo{pages}{1021--1025},
  \doiprefix\url{https://doi.org/10.1016/j.asr.2003.02.031}
  (\bibinfo{year}{2004}).
\newblock \bibinfo{note}{Space Debris}.

\bibitem{Abercromby2005}
\bibinfo{author}{Abercromby, K.}, \bibinfo{author}{Guyote, M.},
  \bibinfo{author}{Okada, J.} \& \bibinfo{author}{et~al.}
\newblock \bibinfo{title}{Applying space weathering models to common spacecraft
  materials to predict spectral signatures}.
\newblock In \emph{\bibinfo{booktitle}{Proceedings of AMOS Technical
  Conference}} (\bibinfo{year}{2005,Maui, Hawaii, USA}).

\bibitem{Reyes2018CharacterizationOS}
\bibinfo{author}{Reyes, J.~A.}, \bibinfo{author}{Cowardin, H.~M.} \&
  \bibinfo{author}{Cone, D.~M.}
\newblock \bibinfo{title}{Characterization of spacecraft materials using
  reflectance spectroscopy}.
\newblock In \emph{\bibinfo{booktitle}{Proceedings of the Advanced Maui Optical
  and Space Surveillance Technologies Conference, Maui, Hawaii, September
  11-14}} (\bibinfo{year}{2018}).

\bibitem{CARDONA2016514}
\bibinfo{author}{Cardona, T.}, \bibinfo{author}{Seitzer, P.},
  \bibinfo{author}{Rossi, A.}, \bibinfo{author}{Piergentili, F.} \&
  \bibinfo{author}{Santoni, F.}
\newblock \bibinfo{journal}{\bibinfo{title}{Bvri photometric observations and
  light-curve analysis of geo objects}}.
\newblock {\emph{\JournalTitle{Advances in Space Research}}}
  \textbf{\bibinfo{volume}{58}}, \bibinfo{pages}{514--527},
  \doiprefix\url{https://doi.org/10.1016/j.asr.2016.05.025}
  (\bibinfo{year}{2016}).

\bibitem{zhao2016multicolor}
\bibinfo{author}{Zhao, X.-F.}, \bibinfo{author}{Zhang, H.-Y.},
  \bibinfo{author}{Yu, Y.} \& \bibinfo{author}{Mao, Y.-D.}
\newblock \bibinfo{journal}{\bibinfo{title}{Multicolor photometry of
  geosynchronous satellites and application on feature recognition}}.
\newblock {\emph{\JournalTitle{Advances in Space Research}}}
  \textbf{\bibinfo{volume}{58}}, \bibinfo{pages}{2269--2279}
  (\bibinfo{year}{2016}).

\bibitem{VANANTI20172488}
\bibinfo{author}{Vananti, A.}, \bibinfo{author}{Schildknecht, T.} \&
  \bibinfo{author}{Krag, H.}
\newblock \bibinfo{journal}{\bibinfo{title}{Reflectance spectroscopy
  characterization of space debris}}.
\newblock {\emph{\JournalTitle{Advances in Space Research}}}
  \textbf{\bibinfo{volume}{59}}, \bibinfo{pages}{2488--2500},
  \doiprefix\url{https://doi.org/10.1016/j.asr.2017.02.033}
  (\bibinfo{year}{2017}).

\bibitem{WILLISON20161318}
\bibinfo{author}{Willison, A.} \& \bibinfo{author}{Bédard, D.}
\newblock \bibinfo{journal}{\bibinfo{title}{A novel approach to modeling
  spacecraft spectral reflectance}}.
\newblock {\emph{\JournalTitle{Advances in Space Research}}}
  \textbf{\bibinfo{volume}{58}}, \bibinfo{pages}{1318--1330},
  \doiprefix\url{https://doi.org/10.1016/j.asr.2016.06.013}
  (\bibinfo{year}{2016}).

\bibitem{Lind2021}
\bibinfo{author}{Lind, L.}, \bibinfo{author}{Laamanen, H.} \&
  \bibinfo{author}{Pölönen, I.}
\newblock \bibinfo{title}{{Hyperspectral imaging of asteroids using an
  FPI-based sensor}}.
\newblock In \bibinfo{editor}{Babu, S.~R.}, \bibinfo{editor}{Hélière, A.} \&
  \bibinfo{editor}{Kimura, T.} (eds.) \emph{\bibinfo{booktitle}{Sensors,
  Systems, and Next-Generation Satellites XXV}}, vol. \bibinfo{volume}{11858},
  \bibinfo{pages}{65 -- 78}, \doiprefix\url{10.1117/12.2599514}.
  \bibinfo{organization}{International Society for Optics and Photonics}
  (\bibinfo{publisher}{SPIE}, \bibinfo{year}{2021}).

\bibitem{Esposito2018}
\bibinfo{author}{Esposito, M.}
\newblock \bibinfo{title}{{Spectral imagers for relative navigation, on-orbit
  servicing and debris removal}}.
\newblock In \emph{\bibinfo{booktitle}{Clean Space Industrial Days}}
  (\bibinfo{publisher}{ESA}, \bibinfo{year}{24 Oct 2018}).

\bibitem{Poojary2015}
\bibinfo{author}{Poojary, N.}, \bibinfo{author}{D'Souza, H.},
  \bibinfo{author}{Puttaswamy, M.~R.} \& \bibinfo{author}{Kumar, G.~H.}
\newblock \bibinfo{title}{Automatic target detection in hyperspectral image
  processing: A review of algorithms}.
\newblock In \emph{\bibinfo{booktitle}{2015 12th International Conference on
  Fuzzy Systems and Knowledge Discovery (FSKD)}}, \bibinfo{pages}{1991--1996},
  \doiprefix\url{10.1109/FSKD.2015.7382255} (\bibinfo{year}{2015}).

\bibitem{wetterer2009}
\bibinfo{author}{C.J., W.} \& \bibinfo{author}{Jah, M.}
\newblock \bibinfo{journal}{\bibinfo{title}{Attitude estimation from light
  curves}}.
\newblock {\emph{\JournalTitle{AIAA Journal of Guidance Control and Dynamics}}}
  \textbf{\bibinfo{volume}{32, No 5}} (\bibinfo{year}{, September-October
  2009}).

\bibitem{dianetti2018}
\bibinfo{author}{Dianetti, A.} \& \bibinfo{author}{Crassidis, J.}
\newblock \bibinfo{title}{Light curve analysis using wavelets}.
\newblock In \emph{\bibinfo{booktitle}{2018 AIAA Guidance, Navigation, and
  Control Conference}} (\bibinfo{year}{8–12 January 2018 Kissimmee,
  Florida}).

\bibitem{Matsushita2019}
\bibinfo{author}{Matsushita, Y.}, \bibinfo{author}{Arakawa, R.},
  \bibinfo{author}{Yoshimura, Y.} \& \bibinfo{author}{Hanada, T.}
\newblock \bibinfo{title}{Light curve analysis and attitude estimation of space
  objects focusing on glint}.
\newblock In \emph{\bibinfo{booktitle}{1st International Orbital Debris
  Conference (IOC)}} (\bibinfo{year}{2019,Sugar Land, Texas, USA.}).

\bibitem{chote2019}
\bibinfo{author}{Chote, P.}, \bibinfo{author}{Blake, J.} \&
  \bibinfo{author}{Pollacco, D.}
\newblock \bibinfo{title}{Precision optical light curves of leo and geo
  objects}.
\newblock In \emph{\bibinfo{booktitle}{The Advanced Maui Optical and Space
  Surveillance Technologies (AMOS) Conference}} (\bibinfo{year}{2019}).

\bibitem{santoni2013}
\bibinfo{author}{Santoni, F.}, \bibinfo{author}{Cordelli, E.} \&
  \bibinfo{author}{Piergentili, F.}
\newblock \bibinfo{journal}{\bibinfo{title}{Determination of
  disposed-upper-stage attitude motion by ground-based optical observations}}.
\newblock {\emph{\JournalTitle{Journal of Spacecraft and Rockets}}}
  \textbf{\bibinfo{volume}{50}}, \bibinfo{pages}{701--708},
  \doiprefix\url{10.2514/1.A32372} (\bibinfo{year}{2013}).
\newblock \eprint{https://doi.org/10.2514/1.A32372}.

\bibitem{YANAGISAWA2012136}
\bibinfo{author}{Yanagisawa, T.} \& \bibinfo{author}{Kurosaki, H.}
\newblock \bibinfo{journal}{\bibinfo{title}{Shape and motion estimate of leo
  debris using light curves}}.
\newblock {\emph{\JournalTitle{Advances in Space Research}}}
  \textbf{\bibinfo{volume}{50}}, \bibinfo{pages}{136--145},
  \doiprefix\url{https://doi.org/10.1016/j.asr.2012.03.021}
  (\bibinfo{year}{2012}).

\bibitem{Kerr2021}
\bibinfo{author}{Kerr, E.} \emph{et~al.}
\newblock \bibinfo{title}{Light curves for geo object characterisation}.
\newblock In \emph{\bibinfo{booktitle}{8th European Conference on Space
  Debris}} (\bibinfo{year}{2021}).

\bibitem{McNally2021}
\bibinfo{author}{McNally, K.}, \bibinfo{author}{Ramirez, D.},
  \bibinfo{author}{Anton, A.}, \bibinfo{author}{Smith, D.} \&
  \bibinfo{author}{Dick, J.}
\newblock \bibinfo{title}{Artificial intelligence for space resident objects
  characterisation with lightcurves}.
\newblock In \emph{\bibinfo{booktitle}{8th European Conference on Space
  Debris}} (\bibinfo{year}{2021}).

\bibitem{vasile2022intelligent}
\bibinfo{author}{Vasile, M.} \emph{et~al.}
\newblock \bibinfo{journal}{\bibinfo{title}{Intelligent characterisation of
  space objects with hyperspectral imaging}}.
\newblock {\emph{\JournalTitle{Acta Astronautica}}}
  \textbf{\bibinfo{volume}{203}}, \bibinfo{pages}{510--534},
  \doiprefix\url{https://doi.org/10.1016/j.actaastro.2022.11.039}
  (\bibinfo{year}{2023}).

\bibitem{vasile2022hyperclass}
\bibinfo{author}{Vasile, M.} \emph{et~al.}
\newblock \bibinfo{title}{Hyperspectral classification of space objects}
  (\bibinfo{year}{2022}).

\bibitem{ZHAO20162269}
\bibinfo{author}{Zhao, X.-F.}, \bibinfo{author}{Zhang, H.-Y.},
  \bibinfo{author}{Yu, Y.} \& \bibinfo{author}{Mao, Y.-D.}
\newblock \bibinfo{journal}{\bibinfo{title}{Multicolor photometry of
  geosynchronous satellites and application on feature recognition}}.
\newblock {\emph{\JournalTitle{Advances in Space Research}}}
  \textbf{\bibinfo{volume}{58}}, \bibinfo{pages}{2269--2279},
  \doiprefix\url{https://doi.org/10.1016/j.asr.2016.09.020}
  (\bibinfo{year}{2016}).

\bibitem{clark2007usgs}
\bibinfo{author}{Clark, R.~N.} \emph{et~al.}
\newblock \bibinfo{title}{Usgs digital spectral library splib06a}.
\newblock \bibinfo{type}{Tech. Rep.}, \bibinfo{institution}{US Geological
  Survey} (\bibinfo{year}{2007}).

\bibitem{winter1999n}
\bibinfo{author}{Winter, M.~E.}
\newblock \bibinfo{title}{N-findr: An algorithm for fast autonomous spectral
  end-member determination in hyperspectral data}.
\newblock In \emph{\bibinfo{booktitle}{Imaging Spectrometry V}}, vol.
  \bibinfo{volume}{3753}, \bibinfo{pages}{266--275}
  (\bibinfo{organization}{SPIE}, \bibinfo{year}{1999}).

\bibitem{harsanyi1994hyperspectral}
\bibinfo{author}{Harsanyi, J.~C.} \& \bibinfo{author}{Chang, C.-I.}
\newblock \bibinfo{journal}{\bibinfo{title}{Hyperspectral image classification
  and dimensionality reduction: An orthogonal subspace projection approach}}.
\newblock {\emph{\JournalTitle{IEEE Transactions on geoscience and remote
  sensing}}} \textbf{\bibinfo{volume}{32}}, \bibinfo{pages}{779--785}
  (\bibinfo{year}{1994}).

\bibitem{nascimento2005vertex}
\bibinfo{author}{Nascimento, J.~M.} \& \bibinfo{author}{Dias, J.~M.}
\newblock \bibinfo{journal}{\bibinfo{title}{Vertex component analysis: A fast
  algorithm to unmix hyperspectral data}}.
\newblock {\emph{\JournalTitle{IEEE transactions on Geoscience and Remote
  Sensing}}} \textbf{\bibinfo{volume}{43}}, \bibinfo{pages}{898--910}
  (\bibinfo{year}{2005}).

\bibitem{heylen2011fully}
\bibinfo{author}{Heylen, R.}, \bibinfo{author}{Burazerovic, D.} \&
  \bibinfo{author}{Scheunders, P.}
\newblock \bibinfo{journal}{\bibinfo{title}{Fully constrained least squares
  spectral unmixing by simplex projection}}.
\newblock {\emph{\JournalTitle{IEEE Transactions on Geoscience and Remote
  Sensing}}} \textbf{\bibinfo{volume}{49}}, \bibinfo{pages}{4112--4122}
  (\bibinfo{year}{2011}).

\bibitem{zhang2016exact}
\bibinfo{author}{Zhang, Z.} \& \bibinfo{author}{Aeron, S.}
\newblock \bibinfo{journal}{\bibinfo{title}{Exact tensor completion using
  t-svd}}.
\newblock {\emph{\JournalTitle{IEEE Transactions on Signal Processing}}}
  \textbf{\bibinfo{volume}{65}}, \bibinfo{pages}{1511--1526}
  (\bibinfo{year}{2016}).

\bibitem{nie2021space}
\bibinfo{author}{Nie, B.}, \bibinfo{author}{Yang, L.}, \bibinfo{author}{Zhao,
  F.}, \bibinfo{author}{Zhou, J.} \& \bibinfo{author}{Jing, J.}
\newblock \bibinfo{journal}{\bibinfo{title}{Space object material
  identification method of hyperspectral imaging based on tucker
  decomposition}}.
\newblock {\emph{\JournalTitle{Advances in Space Research}}}
  \textbf{\bibinfo{volume}{67}}, \bibinfo{pages}{2031--2043}
  (\bibinfo{year}{2021}).

\bibitem{aggarwal2016hyperspectral}
\bibinfo{author}{Aggarwal, H.~K.} \& \bibinfo{author}{Majumdar, A.}
\newblock \bibinfo{journal}{\bibinfo{title}{Hyperspectral unmixing in the
  presence of mixed noise using joint-sparsity and total variation}}.
\newblock {\emph{\JournalTitle{IEEE Journal of Selected Topics in Applied Earth
  Observations and Remote Sensing}}} \textbf{\bibinfo{volume}{9}},
  \bibinfo{pages}{4257--4266} (\bibinfo{year}{2016}).

\bibitem{duzellier2022space}
\bibinfo{author}{Duzellier, S.} \emph{et~al.}
\newblock \bibinfo{journal}{\bibinfo{title}{Space debris generation in geo:
  Space materials testing and evaluation}}.
\newblock {\emph{\JournalTitle{Acta Astronautica}}}
  \textbf{\bibinfo{volume}{192}}, \bibinfo{pages}{258--275}
  (\bibinfo{year}{2022}).

\bibitem{pearce2020examining}
\bibinfo{author}{Pearce, E.~C.}, \bibinfo{author}{Weiner, B.} \&
  \bibinfo{author}{Krantz, H.}
\newblock \bibinfo{journal}{\bibinfo{title}{Examining the effects of on-orbit
  aging of sl-12 rocket bodies using visible band spectra with the mmt
  telescope}}.
\newblock {\emph{\JournalTitle{Journal of Space Safety Engineering}}}
  \textbf{\bibinfo{volume}{7}}, \bibinfo{pages}{376--380}
  (\bibinfo{year}{2020}).

\end{thebibliography}
\section*{Author contributions statement}

M.V. conceived the idea and methodology, directed the implementation of the pipeline and the execution of the experiments, analysed the results and reviewed the paper,  L.W. implemented the simulation model and the machine learning material unmixing, A.C. developed the probabilistic material and object classification, S.M. supported the implementation of the simulation model, V.S. extracted the spectra from the laboratory  experiments, P.M. and S.M. provided key input to the development of the simulations and the analysis of the results.  All authors reviewed the manuscript. 

\section*{Data Availability}
The datasets used and/or analysed during the current study available from the corresponding author on reasonable request.

\section*{Additional information}
 I declare that the authors have no competing interests as defined by Nature Research, or other interests that might be perceived to influence the results and/or discussion reported in this paper.


\end{document}